\documentclass[aps,pra,reprint,superscriptaddress]{revtex4-2}
\usepackage{graphicx}
\usepackage{times}
\usepackage{bm}
\usepackage{amsmath,amssymb}
\usepackage{feynmf}
\usepackage{multirow}
\usepackage[colorlinks=true, allcolors=blue]{hyperref}
\usepackage{xcolor}

\definecolor{RED}{rgb}{1.0,0.0,0.0}

\definecolor{pur}{rgb}{1.0,0.0,1.0}

\begin{document}
\title{Reassessment of line profile asymmetry in measurements of the 1s-2s energy interval in hydrogen}

\author{D. Solovyev} 
\affiliation{ Department of Physics, St. Petersburg State University, Petrodvorets, Oulianovskaya 1, 198504, St. Petersburg, Russia }
\affiliation{ Petersburg Nuclear Physics Institute named by B.P. Konstantinov of National Research Centre 'Kurchatov Institut', St. Petersburg, Gatchina 188300, Russia }
\author{A. Anikin} 
\email[E-mail:]{alexey.anikin.spbu@gmail.com}
\affiliation{ Department of Physics, St. Petersburg State University, Petrodvorets, Oulianovskaya 1, 198504, St. Petersburg, Russia }
\affiliation{ D. I. Mendeleev Institute for Metrology, St. Petersburg, 190005, Russia }
\author{T. Zalialiutdinov}
\affiliation{ Department of Physics, St. Petersburg State University, Petrodvorets, Oulianovskaya 1, 198504, St. Petersburg, Russia }
\affiliation{ Petersburg Nuclear Physics Institute named by B.P. Konstantinov of National Research Centre 'Kurchatov Institut', St. Petersburg, Gatchina 188300, Russia }
\author{L. Labzowsky}
\email{passed away 12.01.2026}
\affiliation{ Department of Physics, St. Petersburg State University, Petrodvorets, Oulianovskaya 1, 198504, St. Petersburg, Russia }
\affiliation{ Petersburg Nuclear Physics Institute named by B.P. Konstantinov of National Research Centre 'Kurchatov Institut', St. Petersburg, Gatchina 188300, Russia }
\date{\today}
\begin{abstract}
Experiments to determine transition frequencies in the hydrogen atom represent some of the most precise spectroscopic measurements and are at a higher level among simple atomic systems. The most persistent measured value in hydrogen is the energy interval corresponding to the $1s-2s$ two-photon transition. The achieved experimental precision is several parts of $10^{-15}$ and has not changed over the last two decades. Although repeated experiments in 2011 and 2013 have improved the accuracy by several times, the frequency value has not changed significantly. On this basis, the frequency of the $1s-2s$ transition holds pivotal for determining physical quantities such as the Rydberg constant and the proton charge radius. Theoretical efforts to study in detail the effects that might influence such precise measurements have not revealed significant contributions. The present work revises the theoretical analysis of the line contour asymmetry and its influence on the determination of the two-photon absorption transition frequency, taking into account the theoretical achievements of recent years in this direction. It is shown that the asymmetry of the observed profile can lead to a $1s-2s$ transition frequency shift at the level of modern experimental accuracy. The found frequency shift is consistent with the line shape model contribution that forms the error budget of the experimental measurements. Adjustment can be carried out on the basis of the asymmetric profile that has become standard in recent years.
\end{abstract}
\maketitle

\section{Introduction}

Measurements of transition frequencies in the hydrogen atom are among the most accurate spectroscopic experiments to date, being second only to optical atomic clocks in accuracy. 
In the experiment \cite{PhysRevLett.84.5496} an unprecedented accuracy of $1.8$ parts at $10^{14}$ for relative magnitude or $\pm 46$ Hz for absolute magnitude of the transition frequency was achieved. A decade later, this experiment was replicated, increasing the accuracy to $10$ Hz \cite{Parthey,Mat}. The observed $1s-2s$ absorption line remains the most accurately measured and yet unchanged, while other frequencies involving one- and two-photon transitions (see for brevity \cite{CODATA-2021} and references therein) are subjected to rigorous theoretical and experimental analyses. Of particular interest in this kind of studies is the inconsistencies in the determination of the proton charge radius, see, e.g., \cite{Pohl,Antognini-2013,H-exp,matveevPRLnew,CARLSON201559,gao2022proton}.

Theoretical analyses carried out during this period \cite{Jent-Mohr,PhysRevA.79.062504,PhysRevA.90.012512,Amaro-2015,Amaro-mH-2015,udem2019_AND} have shown the importance of a detailed description of the absorption process for the precise determination of frequency. Recently, it was demonstrated in \cite{SAZL_PRA2024,SAZL_PhysRevA2024} that the emission and absorption process should be considered in a complete inseparable relation (photon scattering as a whole). The main conclusion of such investigations is that the observed line profile for a particular transition turns out to be distorted. The distortion is stronger the larger is the width of the excited state and the smaller is the energy interval between the resonant and the nearest nonresonant state (allowed according to the atomic transition rules). A significant contribution is caused by the quantum interference effect (QIE) for the states with the same parity \cite{Jent-Mohr}.

The comprehensive analysis of the Ly$_\alpha$ line in hydrogen presented in \cite{Jent-Mohr} was later adapted to the theoretical description of experimental data for the $2s-4p$ absorption spectral line in \cite{H-exp}. An asymmetric line profile (Fano–Voigt) was used to extract the transition frequency, allowing for an accuracy of a few parts in $10^{12}$ relative to the measured one-photon transition energy, see \cite{udem2019_AND} for details. Thus, the treatment of the experimentally observed line contour combined with the $1s-2s$ transition energy has enabled a more accurate determination of the proton charge radius and the Rydberg constant.

It should be noted that the improvement in determining the $1s–2s$ transition frequency does not affect subsequent measurements of the $2s–nd/ns$ frequencies of highly excited states, since the latter have a much larger uncertainty.
The main reason for this is that the excited $2s$-state has a small natural level width of $1.31$ Hz. 
 Theoretical efforts to accurately analyse this particular experiment, see e.g. \cite{LShSP_2007,LShSP_PRL_2007}, have not revealed significant effects on the determination of the transition frequency. In addition, the theoretical description is greatly complicated by the setting of the corresponding measurements \cite{PhysRevLett.84.5496,Parthey,Mat}, in which the excitation and deexcitation regions are separated in space, i.e., the decay of the excited state occurs with a time delay.

Nevertheless, the results provided in \cite{LShSP_2007,LShSP_PRL_2007} are of an evaluative nature. In this paper, we present a more detailed analysis aimed at an accurate theoretical construction of the line profile for the process considered in \cite{PhysRevLett.84.5496,Parthey,Mat}. Given the QIE features that account for the angular correlations between absorbed and emitted photons, the use of an asymmetric line profile, as in \cite{H-exp}, may be important for determining the absorption transition $1s-2s$ frequency. The detection of profile asymmetry, even at the level of the current experimental error, may serve to improve future values of physical constants.

Based on the finite-time quantum electrodynamics theory \cite{FGS91} used in \cite{LShSP_2007,LShSP_PRL_2007}, the line profile  approach \cite{Andr}, and the theory given in \cite{Jent-Mohr}, we consider a two-photon $1s-2s$ excitation followed by a time-delayed one-photon decay to the ground state. The constructed line profile as well as the basic formulae are given in the next part of the paper, and all analytical calculations are provided in the Appendix due to their cumbersome nature. The discussion and conclusions are given in the last part of the paper. Throughout the paper relativistic units are used, where $\hbar=c=m=1$.

\section{One-photon emission profile for the $1s-2s$ two-photon absorption process}

In the experiments \cite{PhysRevLett.84.5496,Parthey,Mat} the emission of photons is stimulated by an external electric field and takes place in a region separated in space from the excitation region (i.e. there exists a flight time during which the atoms excited in the $2s$ state reach the region with the presence of an external electric field), see schematic process in Fig.~\ref{Fig1}. Due to the electric field, states of opposite parity {\cite{Azimov1974,Mohr1978}, in particular $2s$ and $2p$, are mixed, and since the energy level $2p$ has the largest natural width $\sim 10^8$ Hz, even a weak field strength leads to a fast damping. The latter circumstance makes it possible to register absorption by observing Ly$_\alpha$ radiation. By varying the frequency of the incident photons, the emission becomes more intense with exact resonant absorption and hence forms a line profile. This line profile is identified as absorption, although it is detected by an emission process because it is centered around the $1s-2s$ energy interval (see below). As recently shown in \cite{SAZL_PRA2024,SAZL_PhysRevA2024}, this is only relevant within the resonance approximation, and the whole scattering process should be considered.
\begin{figure}[h!]
    \includegraphics[width=0.9\columnwidth]{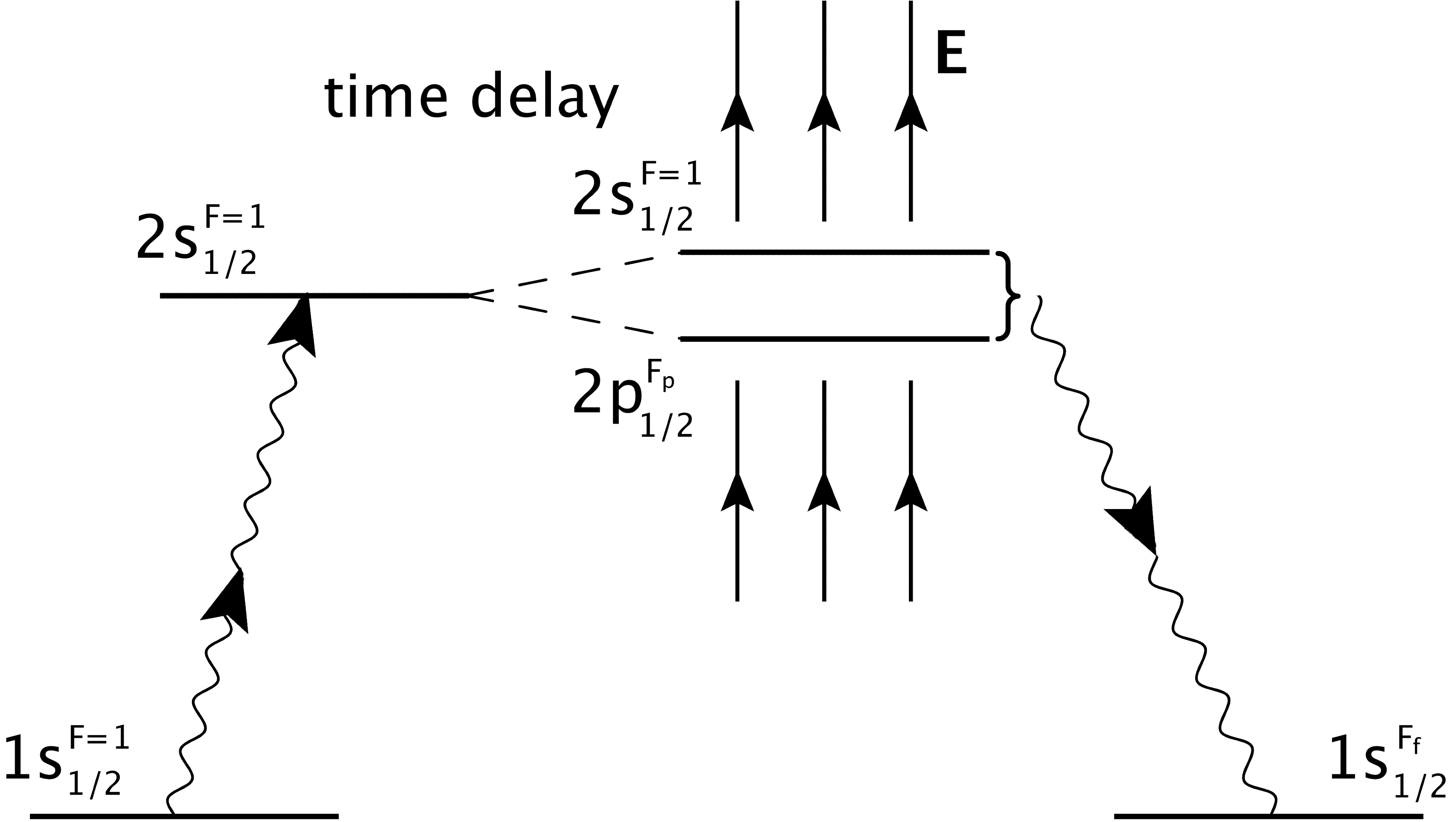}
\caption{
Model scheme of the process used to measure the hydrogen $1s^{F=1}_{1/2} - 2s^{F=1}_{1/2}$ transition frequency. Two photons with the same frequency excite the transition. After a certain delay time, the atoms are exposed to an external electric field $\bm{E}$, which mixes the $2s^{F=1}_{1/2}$ and $2p^{F_p}_{1/2}$, where $F_p = 0,1$, states separated by the Lamb shift. As a result of this mixing, both states contribute to the line profile, leading to its asymmetry. The Lyman-$\alpha$ fluorescence is then observed and measured. For the final state total angular momentum $F_f = 0,1$ and both sublevels contribute to the scattering process.}
 \label{Fig1}
\end{figure}

The starting point for investigating the process of measuring the $1s$–$2s$ energy interval is the electron propagator. It cannot be used in standard form because the eigenfunctions involved in the corresponding decomposition change from unperturbed to solutions in the field. The solution to this issue can be found in \cite{FGS91}, and its application to the process briefly described above was presented in \cite{LShSP_2007,LShSP_PRL_2007}. Omitting details for brevity, the electron propagator operating at finite times can be represented as:
\begin{eqnarray}
\label{1}
S^{\rm FGS}(x_1,x_2) &=& \theta(t_1-t_2)\sum\limits_{\substack{ \tilde{m},n\\ E_{\tilde{m}},E_n>0}}\psi_{\tilde{m}}(x_1)\mathit{w}_{\tilde{m}n} \overline{\psi}_n(x_2) \qquad
\\
\nonumber
&-& \theta(t_2-t_1)\sum\limits_{\substack{ \tilde{m},n\\ E_{\tilde{m}},E_n<0}}\psi_{n}(x_1)\mathit{w}_{n\tilde{m}} \overline{\psi}_{\tilde{m}}(x_2).
\end{eqnarray}

In the expression above $\psi_{\tilde{m}}(x)$ are the solutions of the Dirac equation for the electron in the field of the nucleus and the external electric field, $\psi_n(x)$ are the solutions with zero external field, $\overline{\psi}$ means the Dirac conjugated function; $E_{\tilde{m}},E_n$ are the corresponding eigenvalues and $x={t,\boldsymbol{r}}$ denotes the four-dimensional time-space coordinate. The notation $\theta$ means the Heaviside function. The element $\mathit{w}_{\tilde{m}n}$ in the weak-field approximation reduces to an overlap integral \cite{LSSCK_2009}:
\begin{eqnarray}
\label{2}
\mathit{w}_{\tilde{m}n} = \int d^3r\, \psi^\dagger_{\tilde{m}}(\boldsymbol{r})\psi_n(\boldsymbol{r}).
\end{eqnarray}
The summation in Eq. (\ref{1}) is performed using positive and negative Dirac energies, which is expressed by the corresponding inequalities. 

In further calculations, we neglect the second term in (\ref{1}), referring to the orthogonality property of the wave functions corresponding to the positive and negative energy spectra. In addition, a non-relativistic limit will be used, within which the contribution of the negative part of the spectrum is neglected. Throughout the paper hyperfine structure (HFS) is taken into account, so that arbitrary state $| a \rangle$ is described by following set of quantum numbers $|nljFM \rangle$. The principal quantum number and orbital angular momentum of an electron are denoted by $nl$, $j$ is electronic total angular momentum quantum number, $\boldsymbol{j} = \boldsymbol{l} + \boldsymbol{s}$, where $\boldsymbol{s}$ is electron spin. Total atomic angular momentum $F$ and its projection $M$ are given by $\boldsymbol{F} = \boldsymbol{j} + \boldsymbol{I}$, where $\boldsymbol{I}$ is nuclear spin. For brevity, in the main text of the paper we use a shorthand notations for atomic states, without indicating the hyperfine structure quantum numbers. In experiments \cite{PhysRevLett.84.5496, Parthey, Mat} transition between particular hyperfine components is studied, $1s^{F = 1}_{1/2} \rightarrow 2s^{F = 1}_{1/2}$, so that quantum interference between different HFS pathways does not take place. We also consider only the following set of states: $\tilde{m}\rightarrow 2\tilde{s}_{1/2},2\tilde{p}_{1/2}$ and $n\rightarrow 2s_ {1/2},2p_{1/2}$. The reason for this is the weak electric field of the order of a few tens of V/cm used in the experiment (full mixing occurs in a field of $475$ V/cm \cite{Azimov1974}), and the fact that the closest states are separated by a Lamb shift. Therefore, according to the theory of the quantum interference effect, all other states except $2p_{1/2}$ give a contribution at least an order of magnitude smaller.

The amplitude of the process under consideration can be evaluated within the $S$-matrix formalism, see \cite{LShSP_2007,LShSP_PRL_2007,LSSCK_2009} and Appendix~\ref{supp1} for details. Considering the modified electronic propagator Eq.~(\ref{1}), this yields
\begin{eqnarray}
\label{3}
U_{fi}\sim \frac{A^{(1\gamma)}_{f\,2\tilde{s}_{1/2}} \mathit{w}_{2\tilde{s}_{1/2}2s_{1/2}} A^{(2\gamma)}_{2s_{1/2}\, i}}{E_{2\tilde{s}}-E_f-\omega_f-\mathrm{i} 0}+
\qquad
\\
\nonumber 
 \frac{A^{(1\gamma)}_{f\,2\tilde{s}_{1/2}}\mathit{w}_{2\tilde{s}_{1/2}2p_{1/2}} A^{(2\gamma)}_{2p_{1/2}\, i}}{E_{2\tilde{s}}-E_f-\omega_f-\mathrm{i} 0}
+\frac{ A^{(1\gamma)}_{f\,2\tilde{p}_{1/2}} \mathit{w}_{2\tilde{p}_{1/2}2p_{1/2}} A^{(2\gamma)}_{2p_{1/2}\, i}}{E_{2\tilde{p}}-E_f-\omega_f-\mathrm{i} 0}
\\
\nonumber
+ \frac{A^{(1\gamma)}_{f\,2\tilde{p}_{1/2}}\mathit{w}_{2\tilde{p}_{1/2}2s_{1/2}} A^{(2\gamma)}_{2s_{1/2}\, i}}{E_{2\tilde{p}}-E_f-\omega_f-\mathrm{i} 0}.
\qquad
\end{eqnarray}
Here the final and initial states are denoted as $f$ and $i$ respectively, the frequency of the emitted photon is given by $\omega_f$, the infinitesimal additive in the denominator is left to demonstrate the need to regularize the expression in the case of resonance. 

The one-photon amplitude $A^{(1\gamma)}_{a\, b}$ for the transition $b\rightarrow a$ is given by
\begin{eqnarray}
\label{A_1ph}
A^{(1\gamma)}_{a b} \equiv \langle a|(\boldsymbol{e}_f\boldsymbol{\alpha}) e^{-\mathrm{i}\boldsymbol{k}_f\boldsymbol{r}} |b\rangle,
\end{eqnarray}
where $\boldsymbol{e}_f, \boldsymbol{k}_f$ are the polarization and propagation direction vectors. The two-photon absorption amplitude is defined as
\begin{eqnarray}
\label{A_2ph}
A^{(2\gamma)}_{n\,i} \equiv \sum\limits_h \frac{\langle n| (\boldsymbol{e}_1\boldsymbol{\alpha})e^{\mathrm{i}\boldsymbol{k}_1\boldsymbol{r}} |h\rangle\langle h| (\boldsymbol{e}_2\boldsymbol{\alpha})e^{\mathrm{i}\boldsymbol{k}_2\boldsymbol{r}} |i \rangle}{E_i+\omega-E_h}.
\end{eqnarray}
Indices $1,2$ here correspond to absorbed photons. In general case one should consider permutation of photons. The infinitesimal imaginary part in the energy denominator is omitted in Eq.~(\ref{A_2ph}), which is relevant for the $1s-2s$ transition, see \cite{Akhiezer} and discussion in \cite{LSP_2009,Anikin_2023}. 

It can be seen immediately that the two-photon absorption amplitude is given by different photon multipoles for the states $2s$, $2p$: absorption up to atomic level $2s$ is due to two electric dipole photons E1E1, and excitation of atomic level $2p$ is determined by the sum of two-photon absorption E1E2 (electric dipole and electric quadrupole) and E1M1 (electric and magnetic dipoles) \cite{LSS_2005,SSLP_2010,SS_2015}. Also note that the role of the first and third summands has been considered earlier in \cite{LShSP_2007,LShSP_PRL_2007}, while the other contributions have been omitted.

The smallness of the additional terms relative to the resonant term (the first term in Eq.~\eqref{3}) arises for the following reasons. 
a) Higher multipole photon contributions introduce an additional factor of the fine structure constant, $\alpha$. For instance, the two-photon transition probabilities scale as $W_{2s,1s}^{(2\gamma)} \sim m\alpha^2(\alpha Z)^6$ for the E1E1 process and $W_{2p,1s}^{(2\gamma)} \sim m\alpha^2(\alpha Z)^8$ for the E1E2 + E1M1 processes, where $Z$ is the nuclear charge~\cite{LSS_2005}. 
b) The electric field strength, which enters the tilde wave functions, is also weak. For example, a field of 1 V/cm corresponds to a value of approximately $1.945 \times 10^{-10}$ in atomic units, making its contribution negligible in comparison to the dominant resonant term. However, not all summands in (\ref{3}) turn out to be proportional to the field. 

To trace it accurately, the wave functions in the field are represented as follows \cite{SSLP_2010}:
\begin{eqnarray}
\label{4}
|2\tilde{s},M\rangle = |2s,M\rangle + \sum\limits_{M'}\frac{\langle 2p,M'|e\bm{E}\boldsymbol{r}|2s,M\rangle}{\Delta E_L+\frac{\mathrm{i}}{2}\Gamma_{2p}}|2p,M'\rangle,
\\
\nonumber
|2\tilde{p},M\rangle = |2p,M\rangle - \sum\limits_{M'}\frac{\langle 2s,M'|e\bm{E}\boldsymbol{r}|2p,M\rangle}{\Delta E_L+\frac{\mathrm{i}}{2}\Gamma_{2p}}|2s,M'\rangle.
\end{eqnarray}
The expression (\ref{4}) is written in accordance with perturbation theory, where the dipole interaction with the external electric field $\bm{E}$ is assumed to be small. Although the wave functions of Eq.~(\ref{4}) are not normalized to unity, within the framework of perturbation theory, this manifests as a correction of the next order. The projections $M$ and $M'$ correspond to the total angular momentum (taking into account the hyperfine structure) of the considered state. The natural level width of the $2p$-state $\Gamma_{2p}$ is an order of magnitude smaller than the Lamb shift and has been omitted for the estimates in \cite{LShSP_2007,LShSP_PRL_2007}. However, the presence of level width plays an important role in preserving $T$-invariance, see \cite{Azimov1974}, and, moreover, is essential in determining interference contributions.

Then, taking into account the orthogonality property of wave functions, for the $\mathit{w}_{\tilde{m}n}$ we obtain
\begin{eqnarray}
\label{5}
\mathit{w}_{2\tilde{s}_{1/2}2s_{1/2}} = 1,\,\,\,
\mathit{w}_{2\tilde{s}_{1/2}2p_{1/2}}=\frac{\langle 2s,M|e\bm{E}\boldsymbol{r}|2p,M'\rangle}{\Delta E_L-\frac{\mathrm{i}}{2}\Gamma_{2p}},
\qquad
\\
\nonumber
\mathit{w}_{2\tilde{p}_{1/2}2p_{1/2}}= 1,\,\,\, \mathit{w}_{2\tilde{p}_{1/2}2s_{1/2}} = -\frac{\langle 2p,M|e\bm{E}\boldsymbol{r}|2s,M'\rangle}{\Delta E_L-\frac{\mathrm{i}}{2}\Gamma_{2p}}.
\qquad
\end{eqnarray}
Based on the presented formulae, the line contour determined by radiation can be derived. The details of the derivation are presented in Appendix~\ref{supp2}. 

Using the theory of \cite{Jent-Mohr}, the final result for the line profile can be reduced to the following form:
\begin{eqnarray}
\label{7}
\phi(x) \sim \frac{C}{\left[x-\Delta(x)\right]^2+\frac{1}{4}\Gamma_{2\tilde{s}}^2},
\\
\nonumber
\Delta(x) = \frac{a}{2C}\left(x^2+\frac{1}{4}\Gamma_{2\tilde{s}}^2\right)^2 + \frac{b}{2C}\left(x^2+\frac{1}{4}\Gamma_{2\tilde{s}}^2\right).
\end{eqnarray}
In Eq.~(\ref{7}) a 'mixed' level width of $(2s,2p)$ level is introduced, $\Gamma_{2\tilde{s}}\approx \left(\frac{\bm{E}}{475 \text{V/cm}}\right)^2\Gamma_{2p}$ \cite{SSLP_2010}. In a field of $10$ V/cm, this width is about 44 kHz, which significantly exceeds the width established in \cite{Parthey,Mat}, approximately equal to $1$ kHz. This discrepancy is attributed to the assumption that the field strength is constant and uniform throughout space, instead of being treated as a function $\bm{E}(\boldsymbol{r})$ (where $\boldsymbol{r}$ denotes the spatial coordinate within the de-excitation region, in which non-adiabatic field switching on takes place). The effects of non-adiabatic field switching are discussed further in section~\ref{non-adiab}, see also Appendix~\ref{supp5} for details.
At the same time, the result for the width of $1$ kHz was obtained through the processing of experimental data and subsequent use of the Lorentzian contour. The line profile (\ref{7}) enables similar processing of experimental data, now accounting for asymmetry using parameters $a$, $b$, $C$, and the line width. The following estimates of the frequency shift, including $\Gamma_{2\tilde{s}}$, can be considered as an upper limit. The application of an asymmetric line profile (\ref{7}) for processing experimental data should contribute to further improving the accuracy of determining the frequency of the two-photon $1s-2s$ transition.

The coefficients $a, b, C$ are defined similarly to \cite{Jent-Mohr}. Assuming that the propagation directions of the absorbed photons are anti-collinear, $\boldsymbol{\nu}_1 = - \boldsymbol{\nu}_2$, which corresponds to the experimental setup \cite{PhysRevLett.84.5496, Parthey, Mat}, one can obtain
\begin{eqnarray}
\label{8}
C=|\eta|^2\left(\left|A_1\right|^2 + \left|A_1\right|^2  \Gamma_{2p} \Gamma_{2\tilde{s}} |\eta|^2\right),
\nonumber
\\
a= \left(\frac{\left|A_3\right|^2 }{\Delta E_L^3}+\frac{\left|A_1\right|^2 |\eta|^2 }{\Delta E_L^3}\right),\qquad\qquad
\\
\nonumber
b = 2 \left|A_1\right|^2 \frac{\left|\eta\right|^4}{\Delta E_L}\left(\Delta E_L^2-\frac{\Gamma_{2p}^2}{4}
 + \frac{\Gamma_{2p} \Gamma_{2\tilde{s}}}{2}\right).
\end{eqnarray}
In the above coefficients, the amplitudes are defined as $A_1=A^{(1\gamma)}_{1s\, 2p} \times A^{(ext)}_{2p\, 2s} A^{(2\gamma)}_{2s\,1s}$ and $A_3=A^{(1\gamma)}_{1s\, 2p} A^{(2\gamma)}_{2p\,1s}$, with $A^{(ext)}_{2p\, 2s}\equiv \langle 2p| e\bm{E}\boldsymbol{r}|2s\rangle$ depending on the external field $\bm{E}$.

The comparison of the asymmetric line profile (\ref{7}) with the symmetric Lorentz contour, when $\Delta(x) = 0$, is presented in Fig.~\ref{Fig2}.
\begin{figure}[h!]
    \includegraphics[width=0.9\columnwidth]{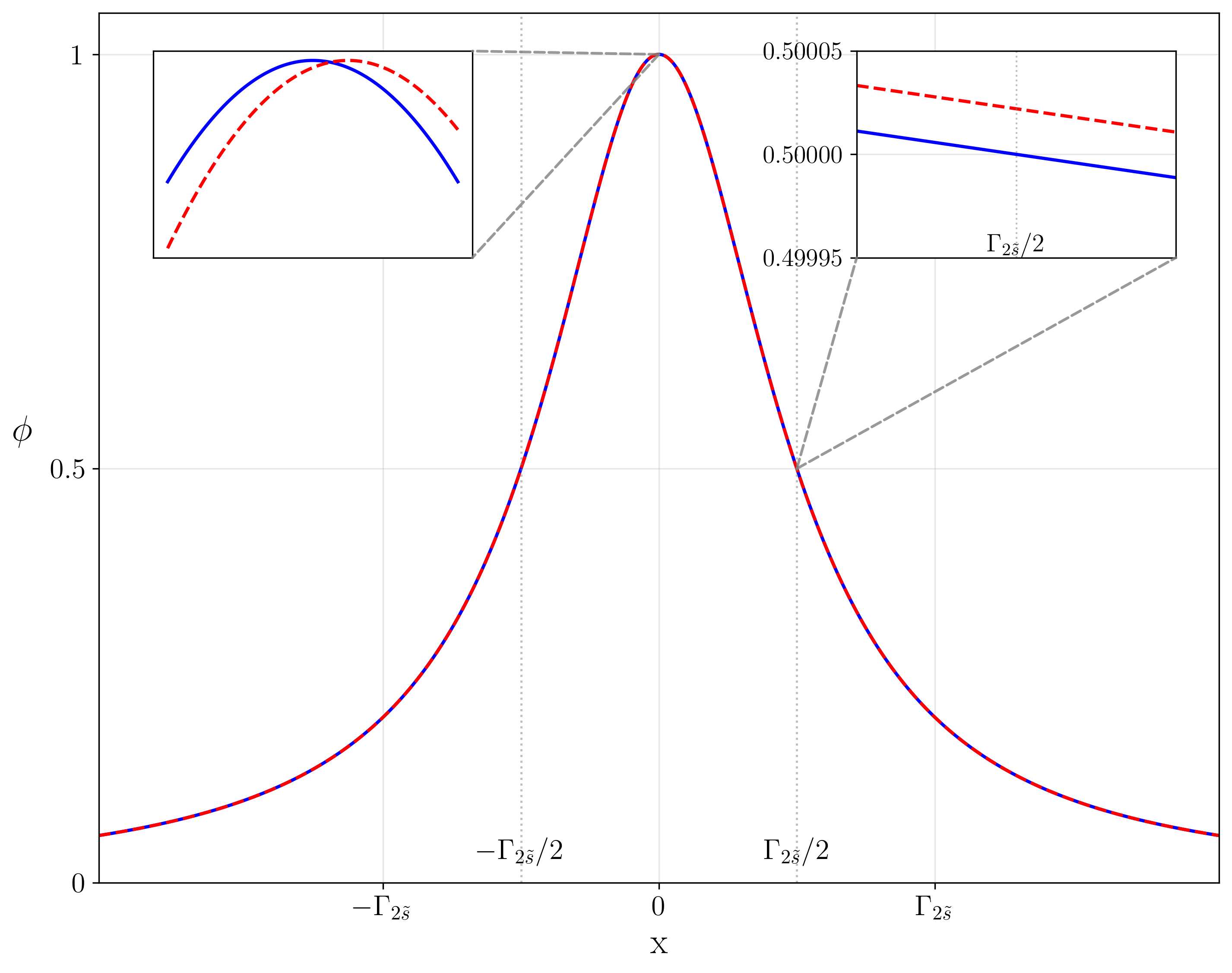}
\caption{
Comparison of the asymmetric profile, Eq. (\ref{7}), (red dashed line) and the Lorentz profile, when $\Delta(x)=0$, (blue solid line). Both profiles are plotted using the 'mixed' level width $\Gamma_{2\tilde{s}}$ for the electric field strength $D = |\boldsymbol{E}| = 10$ V/cm. For better clarity, both functions are normalized to unity. The inserts demonstrate the frequency shifts at the maximum (left insert) and at the full-width half-maximum (right insert), which are indistinctable to the naked eye.
 }
 \label{Fig2}
\end{figure}
The asymmetric Fano contour (\ref{7}) can be used to determine the shift at a given value of frequency \cite{Jent-Mohr}. At the maximum of the line $x=0$ and at the full-width half-maximum $x=\Gamma_{2\tilde{s}}/2$, the frequency shifts are
\begin{eqnarray}
\label{9}
\Delta(0) = \frac{b \Gamma_{2\tilde{s}}^2}{8 C} + \frac{a \Gamma_{2\tilde{s}}^4}{32 C},
\\
\nonumber
\Delta\left(\pm\frac{\Gamma_{2\tilde{s}}}{2}\right) = \frac{b \Gamma_{2\tilde{s}}^2}{4 C} + \frac{a \Gamma_{2\tilde{s}}^4}{16 C}.
\end{eqnarray}

Further evaluation corresponds to the theoretical calculations of the amplitudes and level widths invloved into the coefficients $a,b,C$ in conjunction with the use of the Lamb shift value $\Delta E_L$. The details of such calculations are given in Appendices~\ref{supp2}, \ref{supp3}, where it is shown that in the leading order
\begin{eqnarray}
\label{10}
\frac{b}{2C} = \frac{4\Delta E_L^2-\Gamma_{2p}(\Gamma_{2p}-2\Gamma_{2\tilde{s}})}{4\Delta E_L^3+\Delta E_L\Gamma_{2p}(\Gamma_{2p}+4\Gamma_{2\tilde{s}})}.
\end{eqnarray}
For the $\Delta(x)$ defined at full-width half-maximum of the line profile, it reduces to
\begin{eqnarray}
\label{11}
\Delta\left(\pm \frac{\Gamma_{2\tilde{s}}}{2}\right) = \frac{b\Gamma_{2\tilde{s}}^2}{4C} = [0.9 ;~ 14.8]\text{ Hz},
\end{eqnarray}
where values are given for the field strength $[10;~20]$ V/cm. The shift of the maximum is two times smaller, see Eq.~(\ref{9}). The contribution $a \Gamma_{2\tilde{s}}^4/16 C$ can be excluded from consideration, see Appendix ~\ref{supp4}.

\section{Account for space-separated electric field switching}
\label{non-adiab}

An important part of the line profile analysis for experiments of the type described in \cite{PhysRevLett.84.5496,Parthey,Mat} involves accounting for the spatial separation between the excitation and de-excitation (detection) zones. According to \cite{Kol,PhysRevA.59.1844}, atoms in the excitation region follow a Maxwellian velocity distribution. As a result, the time at which a given atom enters the de-excitation zone
varies from atom to atom. Upon reaching the region where an external electric field is present, the atoms decay immediately, emitting Lyman-$\alpha$ fluorescence. This fluorescence signal is detected after a curtain time delay $\tau$, relative to the moment the excitation laser beam is blocked.
To model this effect, we consider the non-adiabatic switching of the electric field at time $\tau$, which can be treated as a perturbation

\begin{eqnarray}
\label{pert_non-adiab}
V(t) = 
\begin{pmatrix}
   0 & A^{(ext)}_{2p,2s} \\
   A^{(ext)}_{2s,2p} & 0
\end{pmatrix}
\theta(t - \tau).
\end{eqnarray}
By restricting our analysis to the subspace spanned by the $2s_{1/2}$ and $2p_{1/2}$ states, described by the matrix (\ref{pert_non-adiab}), and applying standard time-dependent perturbation theory (detailed in Appendix~\ref{supp5}), we found that
\begin{eqnarray}
\label{width_t}
    \Gamma_{2\tilde{s}}(t) \approx \Gamma_{2s} + 2\left( 1 - \cos \left[ (t - \tau) \Delta E_L \right] \right) \Gamma_{2\tilde{s}}.
    \\
    \nonumber
    C(t) \equiv 2\left( 1 - \cos \left[ (t - \tau) \Delta E_L \right] \right) C.
\end{eqnarray}
This expression incorporates the natural width of the $2s$ state, which was omitted from Eq.~(\ref{7}) due to its smallness. Consequently, the numerator of the line profile Eq.~(\ref{7}) becomes time-dependent, as does the width of the mixed $2\tilde{s}$ state.

Given that atoms traverse the distance $l$ through the de-excitation region with speed $v$, we may substitute $t = l/v$ into Eq.~(\ref{width_t}). To further account for the second-order Doppler effect, we follow Refs. \cite{Sob,riehle} (see Appendix~\ref{supp5} for details). The resulting modified line profile $\phi_{\tau}(x)$ is then given by
\begin{eqnarray}
	\label{line_profile_tau}
	\phi_{\tau} (x) \sim \int\limits_0^{v_{\mathrm{max}}} \frac{v^3 e^{-(v / v_0)^2} \left( 1 - \cos \left[ \left(\frac{l}{v} - \tau\right) \Delta E_L \right] \right) dv}{\left[ x + \omega_0 \frac{v^2}{c^2} \right]^2 + \frac{1}{4}\Gamma^2_{2\tilde{s}}\left(\frac{l}{v}\right)}.\qquad
\end{eqnarray}
The upper limit of integration is chosen as either $v_{\mathrm{max}} = \infty$ or $v_{\mathrm{max}} = l/\tau$, governed by the value of the time delay $\tau$. The quantity $v_0$ appearing in Eq.~(\ref{line_profile_tau}) is given by $v_0 = \sqrt{2kT/M}$, with $M$ denoting the hydrogen atomic mass, $k$ the Boltzmann constant, and $T$ the temperature expressed in Kelvin. 

Taking into consideration the second-order Doppler effect and the space-separated field switching, the expression for the line profile asymmetry shift becomes
\begin{eqnarray}
\label{delta_tau}
\Delta(x, t) \approx \frac{4\Delta E_L^2-\Gamma_{2p}\left[\Gamma_{2p}-2\Gamma_{2\tilde{s}} (t)\right]}{4\Delta E_L^3+\Delta E_L\Gamma_{2p} \left[\Gamma_{2p}+4\Gamma_{2\tilde{s}}(t)\right]}
\\
\nonumber
\times  \left[\left( x + \omega_0 \frac{v^2}{c^2} \right)^2+\frac{1}{4}\Gamma_{2\tilde{s}}(t)^2\right].
\end{eqnarray}
Eq.~(\ref{delta_tau}) contains the leading-order contribution, with the factor $b/2C$ given by Eq.~(\ref{10}), $\omega_0$ is the resonant transition frequency.

To illustrate how non-adiabatic electric field switching and time delay affect the line profile, the corresponding graphs for $\tau = 0$ and $\tau = 1210\,\mu\text{s}$ are presented in Figs.~\ref{Fig3} and~\ref{Fig4}, respectively. To handle the highly oscillatory integrand, we employed the standard Python numerical method \texttt{quad} from the \texttt{scipy} library, with an adaptive integration grid. Figures~\ref{Fig3} and~\ref{Fig4} present the line profiles for $T = 100$ K and $l = 10$ cm, adopting the same $l$ value as in \cite{Kol}.
\begin{figure}[h!]
    \includegraphics[width=0.9\columnwidth]{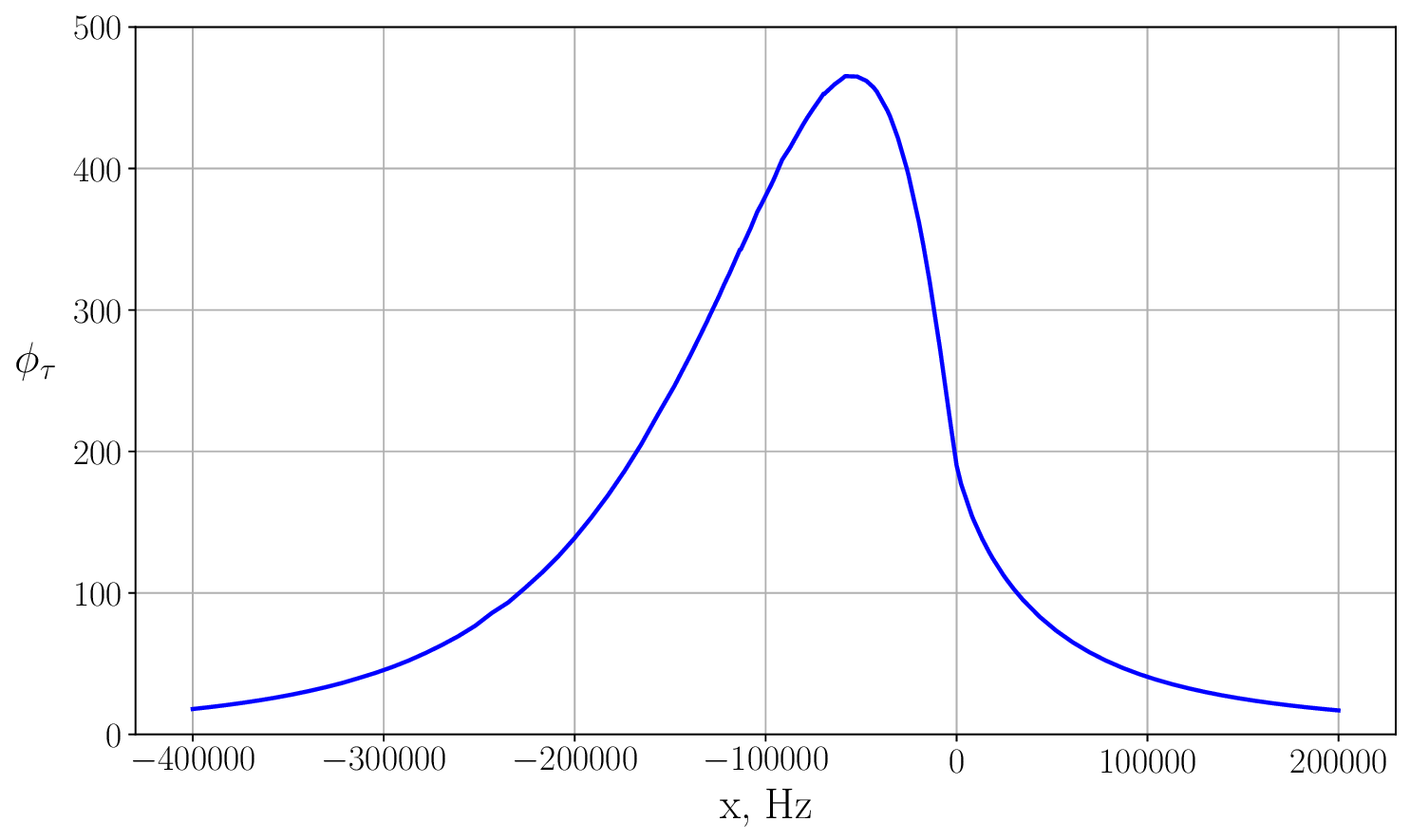}
\caption{The line profile $\phi_{\tau}$, given by Eq.~(\ref{line_profile_tau}), plotted as a function of $x = \omega - \omega_0$ (in Hz) for $T = 100$ K and $v_{\mathrm{max}} = \infty$ (i.e., $\tau = 0$). For illustrative purposes, the function is not normalized. The graph is plotted for the field strength 10 V$/$cm. }
 \label{Fig3}
\end{figure}
\begin{figure}[h!]
	\includegraphics[width=0.9\columnwidth]{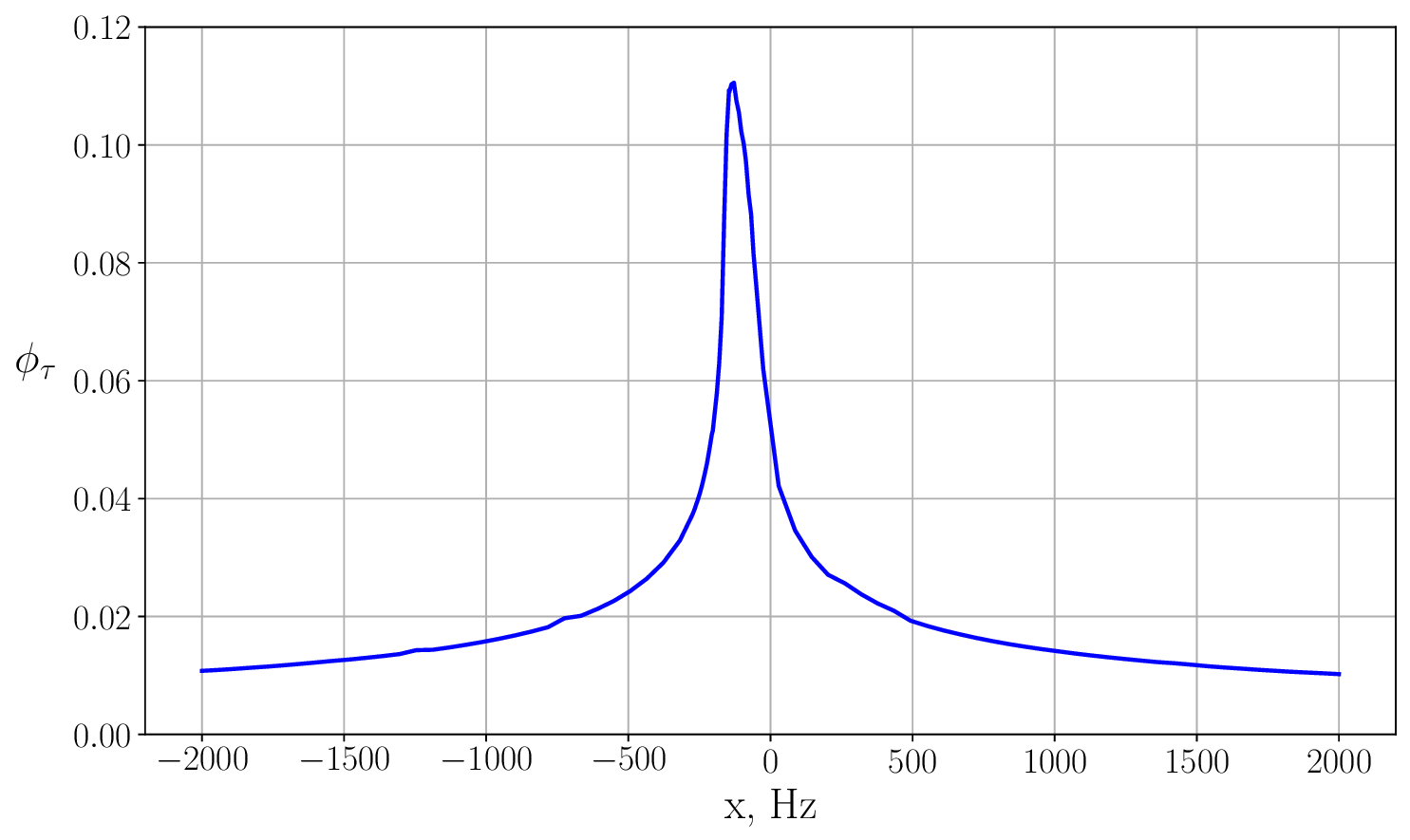}
	\caption{Same as Fig.~\ref{Fig3}, but for the case $v_{\mathrm{max}} = l / \tau \approx 82.645$ m/s, with $l = 10$ cm and $\tau = 1210~\mu$s. }
	\label{Fig4}
\end{figure}

From Fig.~\ref{Fig3}, which corresponds to zero time delay (i.e., fast atoms), one observes that the second-order Doppler effect is significant, resulting in a full width at half maximum (FWHM) on the order of $150$ kHz. In contrast, for a finite time delay $\tau = 1210~\mu$s, which serves to select slower atoms, the line profile (see Fig.~\ref{Fig4}) undergoes a dramatic change. Here, with the atomic velocity constrained by $v_{\mathrm{max}} = l / \tau \approx 82.645$ m/s, both the magnitude of the line profile and the FWHM decrease by orders of magnitude, the latter now being roughly $200$ Hz.

To eliminate the Doppler effect entirely, the atoms in the experiments \cite{PhysRevLett.84.5496,Parthey,Mat,PhysRevA.59.1844} are cooled to about $5$ K. The resulting line profiles are shown in Figs.~\ref{Fig5} and~\ref{Fig6} for $\tau = 0$ and $\tau = 1210~\mu$s, respectively.
\begin{figure}[h!]
    \includegraphics[width=0.9\columnwidth]{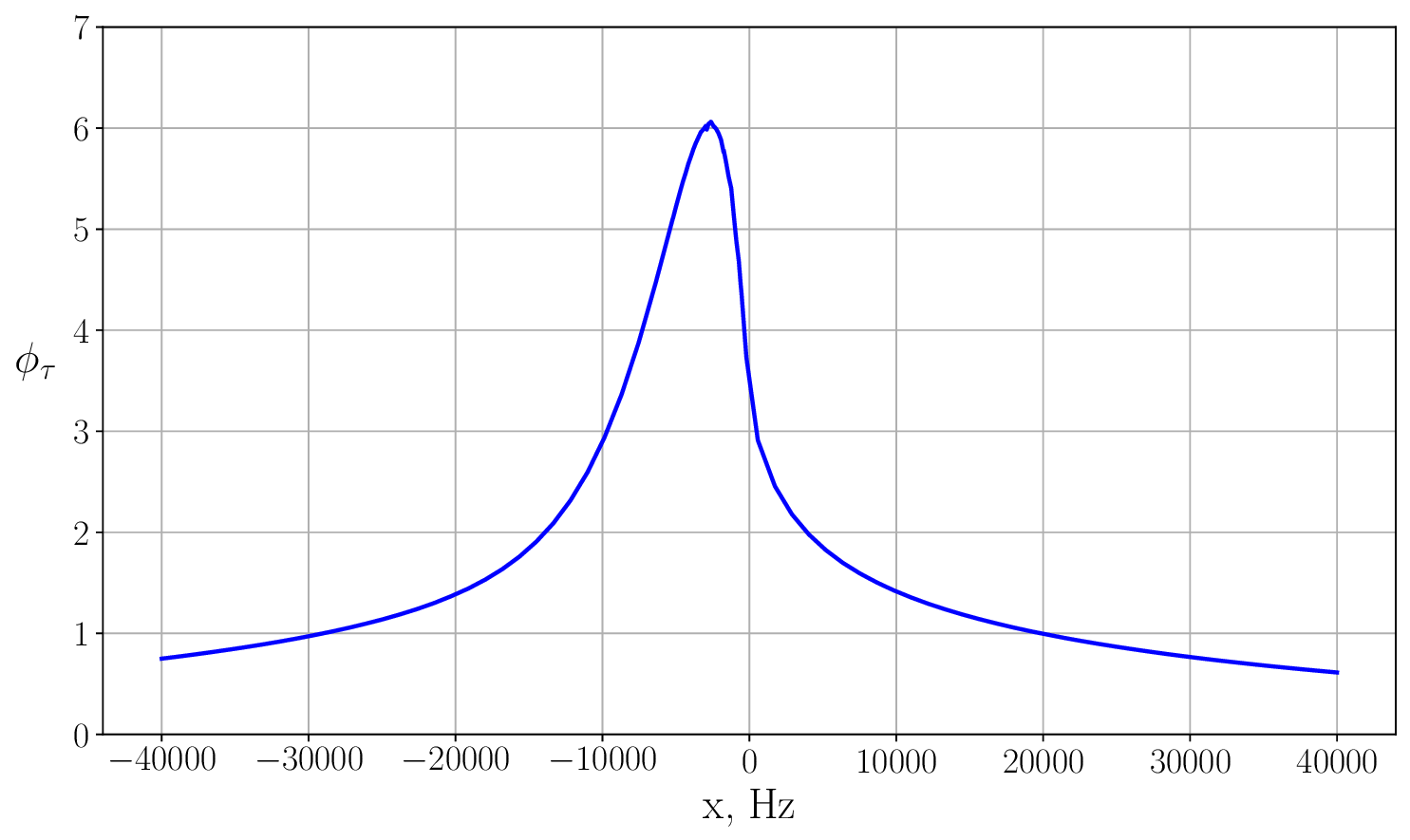}
\caption{The line profile $\phi_{\tau}$, given by Eq.~(\ref{line_profile_tau}), is plotted as a function of $x = \omega - \omega_0$ (in Hz) for $T = 5$ K and $v_{\mathrm{max}} = \infty$ (i.e., $\tau = 0$). The graph is plotted for the field strength 10 V$/$cm}
 \label{Fig5}
\end{figure}
As can be seen from Fig.~\ref{Fig5}, for the cooled atoms without selection using $\tau$, the value of the line profile is significantly lower compared to the $100$ K case. The same can be said about the FWHM, which is roughly $10$ kHz. The corresponding graph for the case when the contribution of slow atoms is sorted out using the time delay is plotted in Fig.~\ref{Fig6}.
 \begin{figure}[h!]
     \includegraphics[width=0.9\columnwidth]{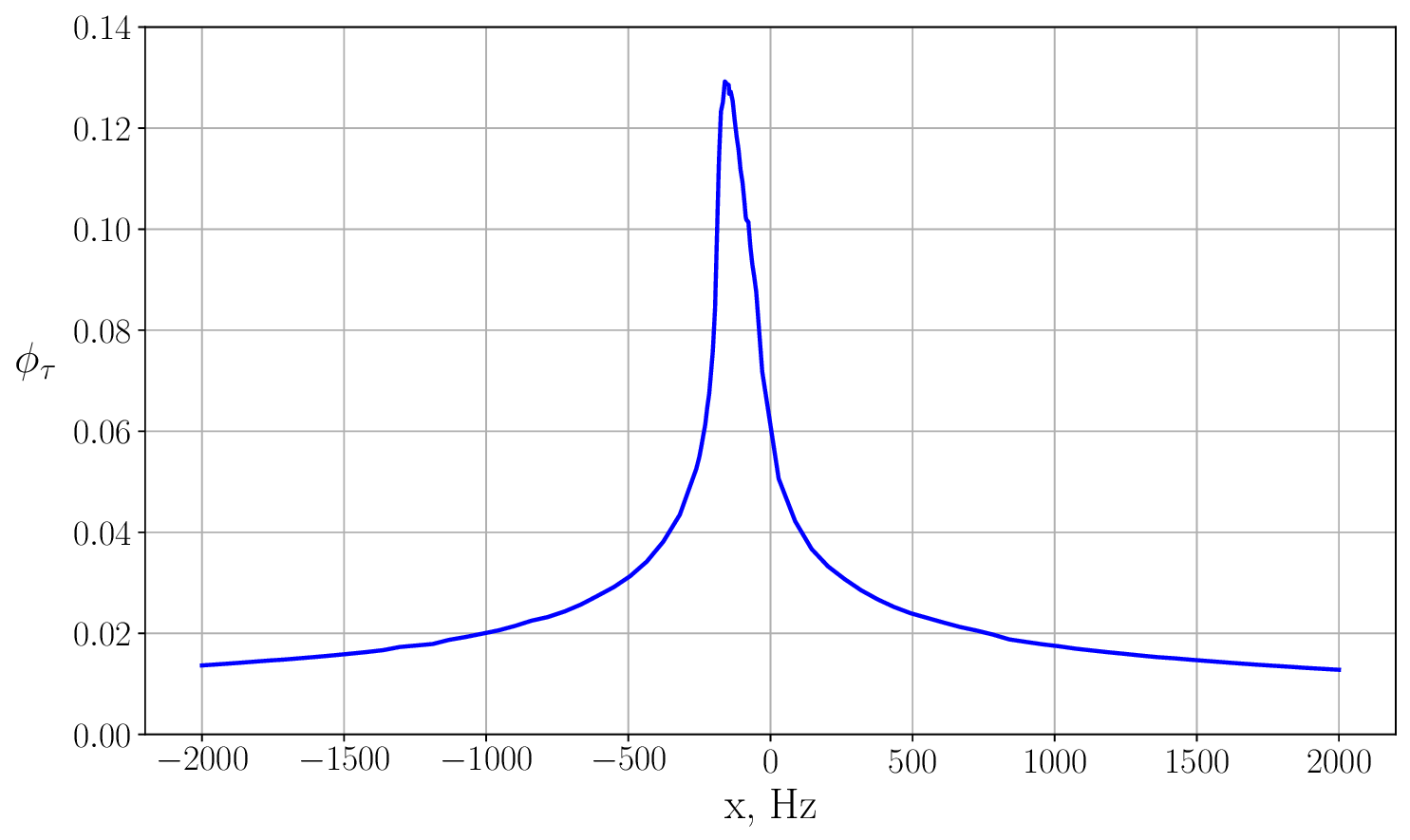}
 \caption{The same as Fig.~\ref{Fig5}, but for the case $v_{\mathrm{max}} = l / \tau \approx 82.645$ m/s, where $l = 10$ cm and $\tau = 1210~\mu$s. }
  \label{Fig6}
 \end{figure}
From Fig.~\ref{Fig6}, the FWHM can be found to be $\approx 200$ Hz. One should mention a relatively large numerical error in the integration in Eq.~(\ref{line_profile_tau}), up to $\lesssim 10$ \% near the peak, due to the highly oscillating parts of the function.

In \cite{Kol}, to derive a line profile for the $1s-2s$ frequency measurement, various broadening effects have been taken into account. As a result, for the case when atoms are cooled to $T = 5$ K and a time delay $\tau = 1210~\mu$s is used, a theoretical result for the linewidth of $550(5)$ Hz was obtained. For the same parameters, the experimental value is $775(20)$ Hz. This $\approx 200$ Hz disagreement can probably be eliminated by accounting for the mixing of $2s_{1/2}$ and $2p_{1/2}$ in the external electric field and its non-adiabatic switching, resulting in the line profile Eq.~(\ref{line_profile_tau}).

Thus, the time delay serves as a regularization parameter for the observed emission line profile, enabling the detection of a Doppler-free signal. The regularization with $\tau$ makes it possible to resolve a narrower line profile. At first glance, the latter would seem to directly imply the vanishing asymmetry of the observed contour. However, this FWHM-cutting procedure does not signify the absence of the asymmetry shift given by Eq.~(\ref{11}); the shift can still manifest through the ensemble of detected emission. The asymmetry shift $\Delta(x, \tau)$ from Eq.~(\ref{delta_tau}) is shown in Fig.~\ref{Fig7} and \ref{Fig8}.

From Fig.~\ref{Fig7} one finds that $\Delta(x, \tau)$ can reach values up to $25$ Hz for large $x$. The periodic pattern is declared by the corresponding dependence in the mixed level width Eq.~(\ref{width_t}).
\begin{figure}[h!]
	\includegraphics[width=0.9\columnwidth]{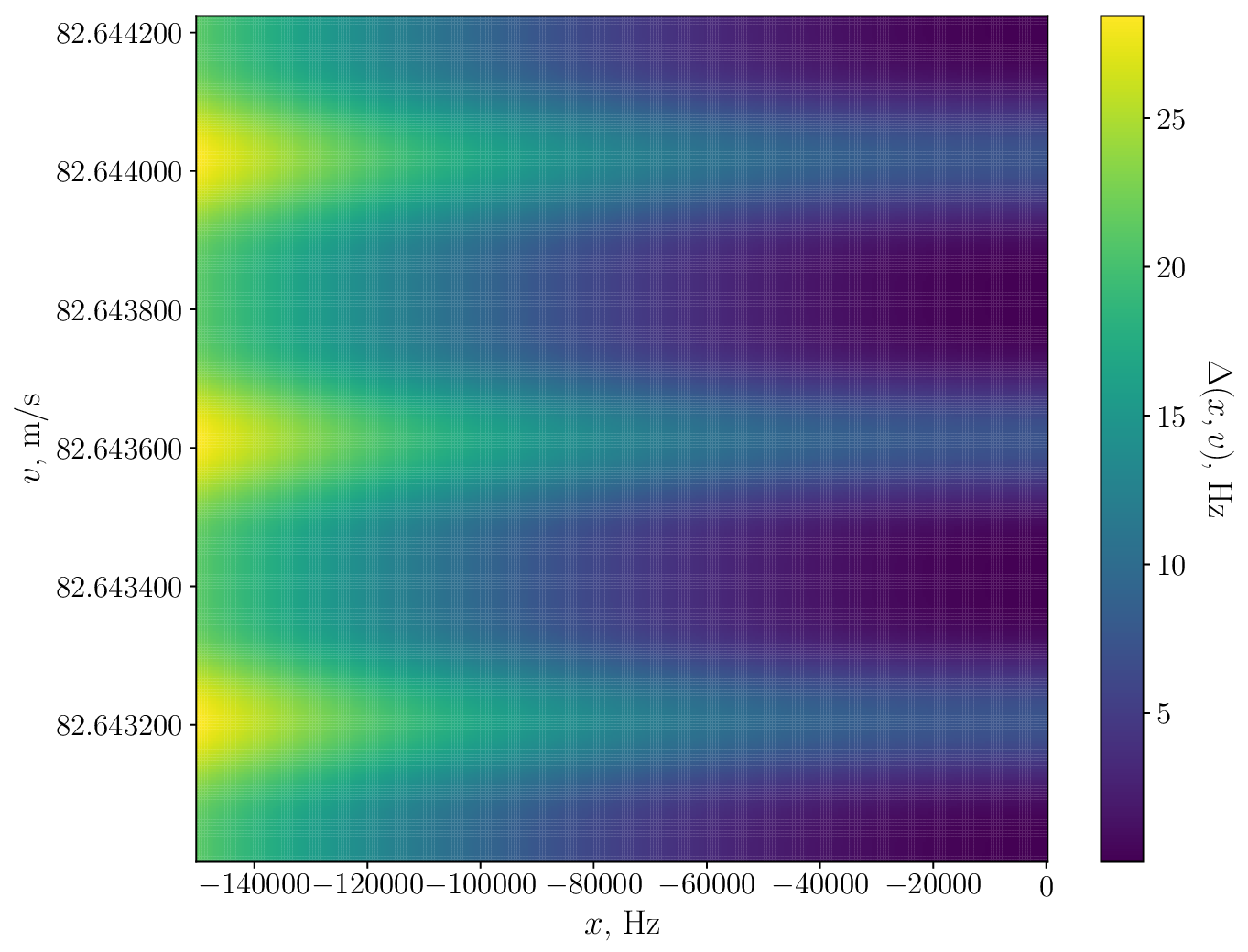}
	\caption{Contour plot for  $\Delta(x, t)$, given by Eq.~(\ref{delta_tau}) at $T = 100$ K, $l = 10$ cm and $\tau = 1210~\mu$s. Color corresponds to the value of the asymmetry shift (Hz), vertical axis illustrates dependence on the velocity of the atoms (m$/$s) and horizontal axis describes frequency dependence (Hz). Limits on the axes are chosen according to the $2\pi$ period of the cosine argument in Eq.~(\ref{width_t}) for the velocity and FWHM of the line profile when $v_{\mathrm{max}} = \infty$ for frequency, see Fig.~\ref{Fig3}. The graph is plotted for the field strength 10 V$/$cm. }
	\label{Fig7}
\end{figure}
The graph in Fig.~\ref{Fig7} is given rather for illustrative purposes. More practical is the $T = 5$ K case, plotted in Fig.~\ref{Fig8}. For this graph, limits for the horizontal axis are chosen according to the value of FWHM from Fig.~\ref{Fig5}.
\begin{figure}[h!]
	\includegraphics[width=0.9\columnwidth]{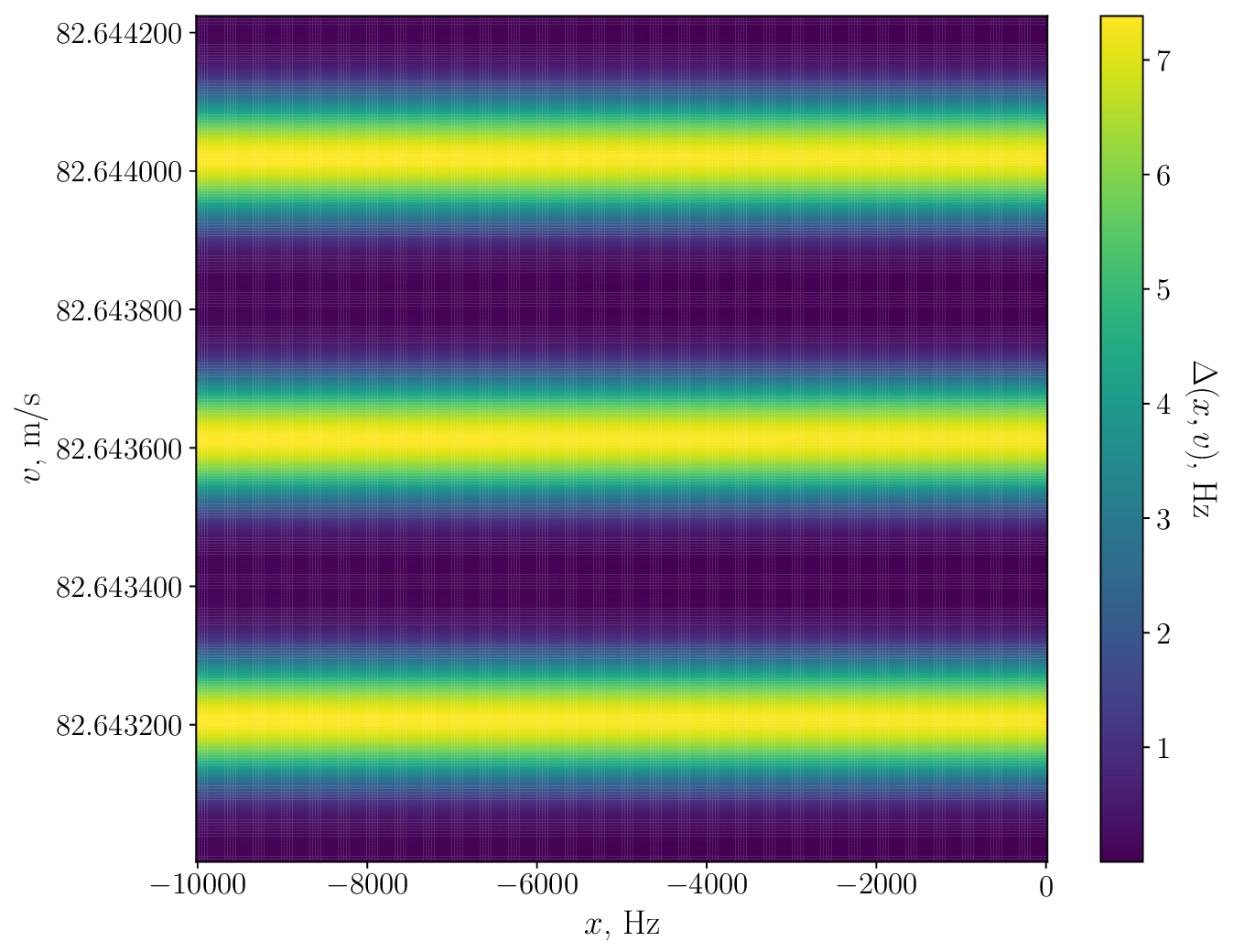}
	\caption{The same as Fig.~\ref{Fig7}, but for the $T = 5$ K case. }
	\label{Fig8}
\end{figure}
As one can see from Fig.~\ref{Fig8}, at a given speed, asymmetry shift depends weakly on frequency. In turn, for particular values of $v$, this shift can reach roughly $7$ Hz value.

\section{Discussion and conclusions}


The above analysis shows that the line profile should be asymmetric. The three quantities $b,C$ and level width included in Eq.~(\ref{7}) can be used as fitting parameters to better match the experimentally observed data. In this approach, the theory can be restricted to a fitting profile (\ref{7}) that effectively accounts for the asymmetry in the linear approximation \cite{H-exp,Jent-Mohr}. The same approach can be used when considering the non-adiabatic field switching and line profile given by Eq.~(\ref{line_profile_tau}). The result obtained earlier in \cite{Jent-Mohr} regarding the off-resonant contribution to the measurement of the $1s-2s$ transition frequency warrants separate discussion. In that work, the authors analyzed quantum interference between transitions to the $2s$ and $ns$ states and found a negligible correction on the order of $10^{-14}$ Hz. This result, however, does not contradict the shift given by Eq.~(\ref{11}). An important distinction is that the influence of an external electric field, and the corresponding mixing of the $2s$ and $2p$ states, was not considered in \cite{Jent-Mohr}. It can therefore be argued that the non-resonant contribution obtained in \cite{Jent-Mohr} and the result of the present work represent separate shifts that, in the general case, are additive.


In contrast to the results of amplitude calculations in photon scattering processes \cite{SZAL_2025}, no angular dependence arises here, see Appendix~\ref{supp3}. The frequency shift Eq.~(\ref{11}, \ref{delta_tau}) does not depend on the electric field orientation, nor on the laser direction and polarization. One should bear in mind the approximate character of the theoretical approach adopted for describing experiments like those in \cite{PhysRevLett.84.5496,Parthey,Mat}. First of all, in the present paper we considered only the external electric field mixing of $2s$ and $2p_{1/2}$ states, see Eq.~(\ref{5}). This is justified by the fact that other states are separated by a much larger energy interval. The nearest sublevel $2p_{3/2}$ is separated by fine structure, which is roughly 10 times larger than the Lamb shift. Stark shift of energy levels was also not taken into account, as in the fields considered it represents a negligible contribution for corrections Eq.~(\ref{9}). The spatial distribution of the electric field was neglected as well. We also considered the case of linear polarizations of the absorbed photons and assumed that there are no effects that change it, see Appendix~\ref{supp3} for details.

At the level of accuracy achieved in \cite{PhysRevLett.84.5496,Parthey,Mat} line profile becomes an observable quantity. Therefore, one has to carefully take into account asymmetry effects and derive the most suitable fitting function for every given experimental set up. The line shape uncertainty, taken into account in \cite{PhysRevLett.84.5496,Parthey,Mat}, in our opinion, can be attributed to the asymmetry effects, discussed in the present paper. In the same time, analysis made explains and justifies from the first principles usage of the asymmetric line profile Eq.~(\ref{7}). Moreover, account of the mixed $2\tilde{s}$ state width and non-adiabatic electric field switching, see Eq.~(\ref{width_t}), can eliminate the $\approx 200$ Hz disagreement between \cite{Kol} and corresponding experimental result for the levelwidth. Analysis made is significantly simplified by the above assumptions, while retaining the possibility of its application to the processing of experimental results similar to \cite{H-exp}. The use of an asymmetric profile should reveal the frequency shift according to the obtained values of the fitting parameters, without necessarily giving estimates of Eq.~(\ref{11}). One can argue that the latter should rather be attributed to the case when electric field is switched on in both interaction and detection regions. When these regions are separated, the line profile Eq.~(\ref{line_profile_tau}) can be used, taking into account the asymmetry shift Eq.~(\ref{delta_tau}). In this case, parameter $\Delta (x , v)$ can still reach values up to the experimental error. Overall, the above theoretical estimates clearly show influence of the studied line profile asymmetry at least on the experimental error and can serve to reduction of transition frequency measurements uncertainty.

{\it Acknowledgments.}
A. A., T. Z. and D. S. acknowledge the support by the Foundation for the Advancement of Theoretical Physics and Mathematics "BASIS"\,  (grants No. 23-1-3-31-1, 22-1-5-9-1 and 25-1-2-18-1).

\bibliographystyle{apsrev4-2}
\bibliography{mybibfile} 


\appendix
\onecolumngrid

\section{Derivation of the two-photon absorption amplitude with delayed decay in an external electric field}
\label{supp1}

To obtain an expression for the amplitude of two-photon absorption followed by a time delayed decay, we start with the ordinary $S$-matrix formalism. A theoretical treatment of this process is presented in \cite{LShSP_2007,LShSP_PRL_2007,LSSCK_2009}. However, here we re-analyse the relevant description to reveal the contributions that may be responsible for the line profile asymmetry for the observed profile matched to the $1s-2s$ two-photon absorption in \cite{PhysRevLett.84.5496,Parthey,Mat}. A Feynman graph representing this process is illustrated in Fig.~\ref{FigS1}.
\begin{figure}[ht]
\label{FigS1}
    \includegraphics[scale=0.25]{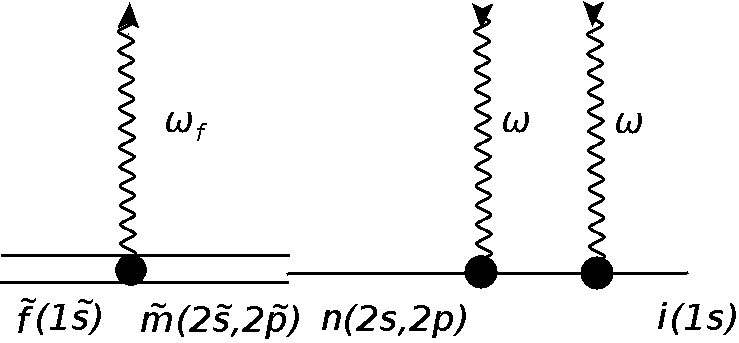}
\caption{
 The process of $1s-2s$ two-photon excitation in the hydrogen atom with subsequent decay in an external electric field. The single solid lines describe the wave functions of the electron and the propagator in the absence of an external electric field. The compound inner electron line represents the electron propagator in the framework of the theory \cite{FGS91}. The outer double solid line corresponds to an atomic electron propagating in an external electric field. As in the standard theory, the wavy lines describe photons. The two absorbed photons are laser photons with frequency $\omega = (E_{2s} - E_{1s})/2$, where $E_n$ are the energies of the atomic electron states in the absence of an external field (eigenstates of the in-Hamiltonian). The emitted photon has frequency $\omega_f$. The designations of states with a tilde ($\tilde{f}=1\tilde{s}\approx 1s$, $\tilde{m}=2\tilde{s}$ or $2\tilde{p}$) correspond to the electronic states in an external field (eigenstates of the out-Hamiltonian).
 }
\end{figure}

Employing the standard Feynman rules, the $S$-matrix element can be written as
\begin{eqnarray}
\label{s1.1}
S_{fi} = (-\mathrm{i}e)^3\int d^4x_1 d^4x_2 d^4x_3 \overline{\psi}_f(x_1)\gamma^\mu A_\mu^*(x_1)
S^{\rm FGS}(x_1,x_2)\gamma^{\nu_1}A_{\nu_1}(x_2)S(x_2,x_3)\gamma^{\nu_2}A_{\nu_2}(x_3)\psi_i(x_3).
\end{eqnarray}
Here $e$ is the electron charge, $x_i={t_i,\boldsymbol{r}_i}$ denotes the four-dimensional time-space coordinate, $\psi_i(x)$ is the initial electron wave function (in the -in space), $\overline{\psi}_f(x)$ is the Dirac-conjugated final state wave function (in the -out space), $\gamma^\mu$ are the standard Dirac matrices, $A_{\mu}(x)$ is the photon wave function with $*$ denoting the emitted photon, $S(x_2,x_3)$ is the ordinary electron propagator in the Furry picture, and $S^{\rm FGS}(x_1,x_2)$ is the electron propagator defined for the QED at finite times \cite{FGS91}.

The expressions for the used quantities can be found as
\begin{eqnarray}
\label{s1.2}
A_{\mu}(x) = \sqrt{\frac{4\pi}{2\omega}} e^{(\lambda)}_{\mu} e^{\mathrm{i}\boldsymbol{k}\boldsymbol{r}-\mathrm{i}\omega t},
\end{eqnarray}
where $e^{(\lambda)}$ represents the polarisation of the photon, $\boldsymbol{k}$ is the wave vector of the photon and $\boldsymbol{r}$ is the spatial vector, $\omega=|\boldsymbol{k}|$ denotes the frequency of the photon and $t$ is the time \cite{Akhiezer}. The eigenmode decomposition of the standard electron propagator for bound states \cite{Andr} is
\begin{eqnarray}
\label{s1.3}
S(x,x') = \frac{\mathrm{i}}{2\pi}\int\limits_{-\infty}^\infty d\Omega\, e^{-\mathrm{i}\Omega(t-t')}\sum\limits_n\frac{\psi_n(\boldsymbol{r})\overline{\psi}_n(\boldsymbol{r'})}{\Omega-E_n(1-\mathrm{i}0)}.\qquad
\end{eqnarray}

Finally, The eigenmode decomposition of FGS propagator \cite{FGS91} reads
\begin{eqnarray}
\label{s1.4}
S^{\rm FGS}(x_1,x_2) = \theta(t_1-t_2)\sum\limits_{\substack{ \tilde{m},n\\ E_{\tilde{m}},E_n>0}}\psi_{\tilde{m}}(x_1)\mathit{w}_{\tilde{m}n} \overline{\psi}_n(x_2) 
- \theta(t_2-t_1)\sum\limits_{\substack{ \tilde{m},n\\ E_{\tilde{m}},E_n<0}}\psi_{n}(x_1)\mathit{w}_{n\tilde{m}} \overline{\psi}_{\tilde{m}}(x_2).
\end{eqnarray}
The first sum in (\ref{s1.4}) describes the propagation of the electron from the spacetime point $x_2$, where there is no additional external field (in-space), to the spacetime point $x_1$, where the field is included (out-space). The eigenfunctions $\psi_n$, $\psi_{\tilde{m}}$ correspond to in- and out-spaces, respectively: $\psi_{\tilde{m}}(x)$ are solutions of the Dirac equation for the electron in the field of the nucleus and the external electric field, $\psi_n(x)$ represent solutions in the absence of the external field, $E_{\tilde{m}}$ and $E_{n}$ are the corresponding eigenvalues. The matrices $\mathit{w}_{\tilde{m}n}$ are defined according to \cite{FGS91} (see also \cite{LSSCK_2009}):
\begin{eqnarray}
\label{s1.5}
\mathit{w}_{\tilde{m}n}=\langle\cdots\tilde{m}\cdots |\hat{\tilde{a}}^\dagger\hat{a}|\cdots n\cdots\rangle,
\end{eqnarray}
where $\langle\cdots\tilde{m}\cdots |$ denotes the out-state vector in Fock space with an electron in state $\tilde{m}$, $|\cdots n\cdots\rangle$ denotes the in-state vector in Fock space with an electron in state $n$, $\hat{\tilde{a}}^\dagger$ is the creation operator in out- Fock space, and $\hat{a}$ is the annihilation operator in in- Fock space. 

In the simple case when electric fields do not create particles, the matrix $\mathit{w}_{\tilde{m}n}$ reduces to the overlap integral:
\begin{eqnarray}
\label{s1.6}
\mathit{w}_{\tilde{m}n}=\int d\boldsymbol{r}\,\psi^\dagger_{\tilde{m}}(\boldsymbol{r})\psi_n(\boldsymbol{r}).
\end{eqnarray}
In the nonrelativistic limit, obviously valid for the neutral hydrogen atom, the Dirac wave functions are replaced by the Schr\"odinger ones and the contribution of negative energies to (\ref{s1.4}) is discarded.

The applicability of the Fradkin-Gitman-Shvartsman picture is defined by the two inequalities \cite{LSSCK_2009}: $\tau_{\rm at}\ll \tau_{\rm field} \ll \tau_{\rm d}$. Here $\tau_{\rm field}$ is the time showing how fast the field is changing in the rest frame of atom, $\tau_{\rm at}$ is the characteristic atomic time necessary for the formation of the stationary atomic states, and $\tau_{\rm d}$ is the atomic decay time. The inequality $\tau_{\rm at}\ll \tau_{\rm field}$ means that the field is changing slowly enough not to destroy the stationary atomic states. The other inequality $\tau_{\rm field} \ll \tau_{\rm d}$ implies that the field is changing quite abruptly in space so that the detection signal has a sharp peak structure which allows for the accurate determinations of the frequency. In \cite{LSSCK_2009} it was calculated that under the conditions of the experiments \cite{PhysRevLett.84.5496,Parthey,Mat} the above inequalities are fulfilled (and we do not repeat the corresponding estimates here for brevity).

To perform integration over the time variables it is convenient to use the Heaviside $\theta$-function representation:
\begin{eqnarray}
\label{s1.7}
\theta(z)=\frac{\mathrm{i}}{2\pi}\lim\limits_{\varepsilon\rightarrow 0^+}\int\limits_{-\infty}^\infty d\xi\frac{e^{-\mathrm{i}z\xi}}{\xi+\mathrm{i}\varepsilon}.
\end{eqnarray}
Integrating over the variables $t_3, t_2$ and $t_1$ in equation (\ref{s1.1}), the product of the Dirac $\delta$-functions can be found: $\delta(E_f+\omega_f-\xi-E_{\tilde{m}})\delta(\xi-\omega+E_n-\Omega)\delta (\Omega-\omega-E_i)$. Here we assume that $E_f$ is weakly affected by the external field (hereafter it is assumed to be equal to the ground state) and that two photons with the same frequencies $\omega$ are used for absorption (the frequency of the emitted photon is denoted by $\omega_f$). Then the remaining integrals over $\xi$ and $\Omega$ are easily evaluated, leading to the expression:
\begin{eqnarray}
\label{s1.8}
S_{fi} = -2\pi\mathrm{i} e^3 \sum\limits_{\tilde{m},n}\delta(E_f+\omega_f-E_{\tilde{m}}+E_n-2\omega-E_i) \frac{(2\pi)^{3/2}}{\omega\sqrt{\omega_f}} N\times \qquad
\\
\nonumber
 \int d\boldsymbol{r}_1 d\boldsymbol{r}_2 d\boldsymbol{r}_3 \overline{\psi}_f(\boldsymbol{r}_1) (\boldsymbol{e}_f\boldsymbol{\alpha}_1) e^{-\mathrm{i}\boldsymbol{k}_f\boldsymbol{r}_1} \frac{\psi_{\tilde{m}}(\boldsymbol{r}_1) \mathit{w}_{\tilde{m}n}  \overline{\psi}_n(\boldsymbol{r}_2)}{E_f+\omega_f-E_{\tilde{m}}(1-\mathrm{i}0)}
(\boldsymbol{e}_1\boldsymbol{\alpha}_2)e^{\mathrm{i}\boldsymbol{k}_1\boldsymbol{r}_2}\sum\limits_h\frac{\psi_h(\boldsymbol{r}_2)\overline{\psi}_h(\boldsymbol{r}_3)}{E_i+\omega-E_h(1-\mathrm{i}0)} (\boldsymbol{e}_2\boldsymbol{\alpha}_3)e^{\mathrm{i}\boldsymbol{k}_2\boldsymbol{r}_3}\psi_i(\boldsymbol{r}_3),
\end{eqnarray}
where $N=(2\pi)^{3/2}/(\omega\sqrt{\omega_f})$ gives the normalization factor arising from the photon wave functions.

The absorbed transverse photons are denoted by indices $1,2$ and the emitted photon by index $f$, $\boldsymbol{\alpha}$ is the Dirac matrix representing the transverse part of $\gamma^\mu$ and denoted according to the integration indices. In principle, a summand with rearranged absorbed photons should be added to the $S$-matrix element. For brevity, we do not illustrate the corresponding contribution since its calculation completely repeats the given derivations. It is worth noting that the expression (\ref{s1.8}) in the resonance approximation can be divided into two parts: a) from $\overline{\psi}_f$ to $\psi_{\tilde{m}}(\boldsymbol{r}_1)$ corresponds to the emission process, and b) the next part represents the two-photon absorption. The first energy denominator is responsible for the formation of the line profile. According to the energy conservation law given by the $\delta$-function in Eq. (\ref{s1.8}), this denominator can be attributed to the absorption process. However, we prefer to keep such denominator, since it is the emitted photon that is registered in the experiment.

Herewith we use the following approximation. First, the summation in Eq. (\ref{s1.8}) includes all possible states from the positive part of the Dirac spectrum (also continuum states), but in the weak-field limit and the resonance approximation all contributions, except for the set of states $\tilde{m}={2\tilde{s},2\tilde{p}}$ and $n={2s,2p}$, are rather small. Second, we neglect the Stark shifts of energy levels $E_{\tilde{m}}$, which is also fulfilled in the weak field limit. Third, since in the framework of Dirac theory the hydrogen states $2s_{1/2}$ and $2p_{1/2}$ remain degenerate (their difference, i.e., the Lamb shift, arises rigorously in the framework of QED theory), we neglect the difference $E_{\tilde{m}}-E_n$ in the argument of the $\delta$-function. As a consequence, we will further consider only pairs of states $\tilde{m}={2\tilde{s}_{1/2},2\tilde{p}_{1/2}}$ and $n= {2s_{1/2},2p_{1/2}}$, but taking into account the hyperfine structure of the states.

Now the eigenfunction of the Hamiltonian taking into account the external electric field can be written in accordance with the perturbation theory \cite{Azimov1974,Mohr1978}:
\begin{eqnarray}
\label{s1.9}
|2\tilde{s},M\rangle = |2s,M\rangle + \sum\limits_{M'}\frac{\langle 2p,M'|e\bm{E}\boldsymbol{r}|2s,M\rangle}{\Delta E_L+\frac{\mathrm{i}}{2}\Gamma_{2p}}|2p,M'\rangle,\qquad
\\
\nonumber
|2\tilde{p},M\rangle = |2p,M\rangle - \sum\limits_{M'}\frac{\langle 2s,M'|e\bm{E}\boldsymbol{r}|2p,M\rangle}{\Delta E_L+\frac{\mathrm{i}}{2}\Gamma_{2p}}|2s,M'\rangle.\qquad
\end{eqnarray}
The expressions (\ref{s1.9}) lead to a small deviation from unity in the normalization factor of wave functions for mixed states. However, being a common coefficient in the amplitude (scattering cross-section), the corresponding correction can be discarded due to multiplication of such a deviation by contributions responsible for the asymmetry of the line profile. A brief discussion of these expressions is given in the main text of the paper. Here $\bm{E}$ represents the electric field strength and the perturbation is taken into account as a dipole interaction, $\Delta E_L$ is the Lamb shift, and $\Gamma_{2p}$ is the natural width of the $2p$ level. The projections $M$ and $M'$ correspond to the total atomic momentum $\boldsymbol{F}=\boldsymbol{j}+\boldsymbol{I}$ ($\boldsymbol{j}$ is the total angular momentum, which is the sum of the orbital momentum $\boldsymbol{l}$ and spin $\boldsymbol{s}$ of the electron, $\boldsymbol{I}$ is the spin momentum of the nucleus).

Then, from Eq. (\ref{s1.6}), we find
\begin{eqnarray}
\label{s1.10}
\mathit{w}_{2\tilde{s}_{1/2}2s_{1/2}} = 1,\,\,\,
\mathit{w}_{2\tilde{s}_{1/2}2p_{1/2}}=\frac{\langle 2s,M|e\bm{E}\boldsymbol{r}|2p,M'\rangle}{\Delta E_L-\frac{\mathrm{i}}{2}\Gamma_{2p}},\,\,\,
\qquad
\\
\nonumber
\mathit{w}_{2\tilde{p}_{1/2}2p_{1/2}}= 1,\,\,\, \mathit{w}_{2\tilde{p}_{1/2}2s_{1/2}} = -\frac{\langle 2p,M|e\bm{E}\boldsymbol{r}|2s,M'\rangle}{\Delta E_L-\frac{\mathrm{i}}{2}\Gamma_{2p}}.
\qquad
\end{eqnarray}
It is also convenient to introduce the following notations:
\begin{eqnarray}
\label{s1.11}
\int d\boldsymbol{r}_1 \overline{\psi}_f(\boldsymbol{r}_1) (\boldsymbol{e}_f\boldsymbol{\alpha}_1) e^{-\mathrm{i}\boldsymbol{k}_f\boldsymbol{r}_1} \psi_{\tilde{m}}(\boldsymbol{r}_1) 
= \langle f|(\boldsymbol{e}_f\boldsymbol{\alpha}) e^{-\mathrm{i}\boldsymbol{k}_f\boldsymbol{r}} |\tilde{m}\rangle
&\equiv& A^{(1\gamma)}_{f\,\tilde{m}},\qquad
\\
\nonumber
\sum\limits_h \frac{\langle n| (\boldsymbol{e}_1\boldsymbol{\alpha})e^{\mathrm{i}\boldsymbol{k}_1\boldsymbol{r}} |h\rangle\langle h| (\boldsymbol{e}_2\boldsymbol{\alpha})e^{\mathrm{i}\boldsymbol{k}_2\boldsymbol{r}} |i \rangle}{E_i+\omega-E_h}
&\equiv& A^{(2\gamma)}_{n\,i},
\end{eqnarray}
where the infinitesimal imaginary part in the energy denominator is omitted, which is relevant for the $1s-2s$ transition, see \cite{Akhiezer} and discussion in \cite{LSP_2009}. In the present work we also neglect the frequency-dependence of the two-photon amplitude $A^{(2\gamma)}_{n\,i}$ due to the smallness of the corresponding asymmetry corrections \cite{Anikin_2023}.

In the resonant approximation taking into account the contribution of the states $\tilde{m}={2\tilde{s},2\tilde{p}}$ and $n={2s,2p}$, four terms can be found:
\begin{eqnarray}
\label{s1.12}
S_{fi} \approx -2\pi\mathrm{i} e^3 \delta(E_f+\omega_f-2\omega-E_i) N \times
\qquad\qquad
\\
\nonumber
\Bigg[ \frac{A^{(1\gamma)}_{f\,2\tilde{s}_{1/2}} \mathit{w}_{2\tilde{s}_{1/2}2s_{1/2}} A^{(2\gamma)}_{2s_{1/2}\, i}}{E_f+\omega_f-E_{2\tilde{s}}(1-\mathrm{i} 0)} + 
\frac{A^{(1\gamma)}_{f\,2\tilde{s}_{1/2}}\mathit{w}_{2\tilde{s}_{1/2}2p_{1/2}} A^{(2\gamma)}_{2p_{1/2}\, i}}{E_f+\omega_f-E_{2\tilde{s}}(1-\mathrm{i} 0)} 
+ \frac{ A^{(1\gamma)}_{f\,2\tilde{p}_{1/2}} \mathit{w}_{2\tilde{p}_{1/2}2p_{1/2}} A^{(2\gamma)}_{2p_{1/2}\, i}}{E_f+\omega_f-E_{2\tilde{p}}(1-\mathrm{i} 0)}
+ \frac{A^{(1\gamma)}_{f\,2\tilde{p}_{1/2}}\mathit{w}_{2\tilde{p}_{1/2}2s_{1/2}} A^{(2\gamma)}_{2s_{1/2}\, i}}{E_f+\omega_f-E_{2\tilde{p}}(1-\mathrm{i} 0)} \Bigg].
\end{eqnarray}
The amplitude of the process under consideration can be obtained via the ordinary relation \cite{Akhiezer}, i.e.
\begin{eqnarray}
\label{s1.13}
S_{fi}= -2\pi\mathrm{i}\delta\left(E_f+\omega_f-2\omega-E_i\right)U_{fi},
\end{eqnarray}
and the differential scattering cross section as
\begin{eqnarray}
\label{s1.14}
d\sigma_{fi}= 2\pi\delta(E_f+\omega_f-2\omega-E_i)\left|U_{fi}\right|^2\mathcal{V},
\end{eqnarray}
where $\mathcal{V}$ denotes the phase volumes for each photon, $\frac{d\boldsymbol{k}_1}{(2\pi)^3}\frac{d\boldsymbol{k}_2}{(2\pi)^3}\frac{d\boldsymbol{k}_f}{(2\pi)^3}$ and should take into account that $\omega_1(\equiv|\boldsymbol{k}_1|)=\omega_2(\equiv|\boldsymbol{k}_2|)$.
 
Introducing notation $\eta = (\Delta E_L+\frac{\mathrm{i}}{2}\Gamma_{2p})^{-1}$ and employing results of Eqs. (\ref{s1.9}), (\ref{s1.10}), the photon scattering amplitude can be reduced to
\begin{eqnarray}
\label{s1.15}
U_{fi} = -e^3N\left[
 \frac{\langle f| (\boldsymbol{e}_f\boldsymbol{\alpha}) e^{-\mathrm{i}\boldsymbol{k}_f\boldsymbol{r}} | 2p\rangle \eta \langle 2p| e\bm{E}\boldsymbol{r}|2s\rangle  A^{(2\gamma)}_{2s\,i}}{E_{2\tilde{s}}-E_f-\omega_f-\mathrm{i} 0}
+  \frac{\langle f| (\boldsymbol{e}_f\boldsymbol{\alpha}) e^{-\mathrm{i}\boldsymbol{k}_f\boldsymbol{r}} | 2p\rangle |\eta|^2 \langle 2p| e\bm{E}\boldsymbol{r}|2s\rangle \langle 2s| e\bm{E}\boldsymbol{r}|2p\rangle  A^{(2\gamma)}_{2p\,i}}{E_{2\tilde{s}}-E_f-\omega_f-\mathrm{i} 0}
 \right.
 \\
 \nonumber
 \left.
+ \frac{\langle f| (\boldsymbol{e}_f\boldsymbol{\alpha}) e^{-\mathrm{i}\boldsymbol{k}_f\boldsymbol{r}} | 2p\rangle   A^{(2\gamma)}_{2p\,i}}{E_{2\tilde{p}}-E_f-\omega_f-\mathrm{i} 0} 
- \frac{\langle f| (\boldsymbol{e}_f\boldsymbol{\alpha}) e^{-\mathrm{i}\boldsymbol{k}_f\boldsymbol{r}} | 2p\rangle \eta^* \langle 2p| e\bm{E}\boldsymbol{r}|2s\rangle  A^{(2\gamma)}_{2s\,i}}{E_{2\tilde{p}}-E_f-\omega_f-\mathrm{i} 0}
\right].
\end{eqnarray}
Here in the notation of the states we omitted the projections of angular momenta for brevity. We also have left the abbreviations for the energies in the field (although the corresponding Stark shift was assumed to be negligibly small) to make it clear that the denominators of the resonance energy are regularized not by natural widths but by widths determined in the presence of an external electric field, see e.g. \cite{SSLP_2010,SS_2015}. The terms related to the one-photon $1s-2s$ amplitudes are discarded in Eq.~(\ref{s1.15}) due to their smallness \cite{LSSPS_2003,LSSCK_2009,SLS_2009,SS_2015}.

It should be noted again that the experiments actually measure the emission profile, not the absorption. Taking into account that the natural width of the $2s$ level is eight orders of magnitude smaller than the width of the $2p$ state, the admixture of the latter due to the external field has a crucial role. In particular, on the basis of the theory of \cite{Azimov1974,Mohr1978} it can be established that the complete mixing of the states $2s$ and $2p$ occurs in a field of $475$ V/cm. In this field the radiation level width of the $2\tilde{s}$ state becomes equal to the level width of the Ly$_\alpha$ line (in turn, the admixture of the $2s$ state to the $2p$ level can be neglected \cite{SSLP_2010}). It has also been shown that the total level width (integrated over all angles and frequencies) depends quadratically on the field strength \cite{SSLP_2010,SS_2015}. Thus, rough estimates of $\Gamma_{2\tilde{s}}$ for the experimental values of the field strength ($D = |E| = [10;~20]$ V/cm) can be given using the factor $\left(D/475\right)^2$, resulting in $\Gamma_{2\tilde{s}}\sim [44;~177]$ kHz. This rough estimate differs significantly from the experimental value of $\Gamma_{\rm exp}\sim 10^3$ Hz. Such a difference can be attributed to the fact that we do not consider the field strength as a function of $\bm{E}(\boldsymbol{x})$ ($\boldsymbol{x}$ is the spatial coordinate of the de-excitation region, where non-adiabatic field switching is possible). A simplest case of non-adiabatic field switching is considered in Appendix~\ref{supp5}.

The line profile for the process under consideration arises from the amplitude Eq. (\ref{s1.15}). Within the resonance approximation the first two terms only can be left. The divergent energy denominator should be regularized. This can be performed with the use of theory \cite{Low}, see also \cite{Andr}. According to this approach the subsequent insertion of the one-loop self-energy correction should be considered, see Fig.~\ref{FigS2}.
\begin{figure}[ht!]\label{FigS2}
    \includegraphics[scale=0.25]{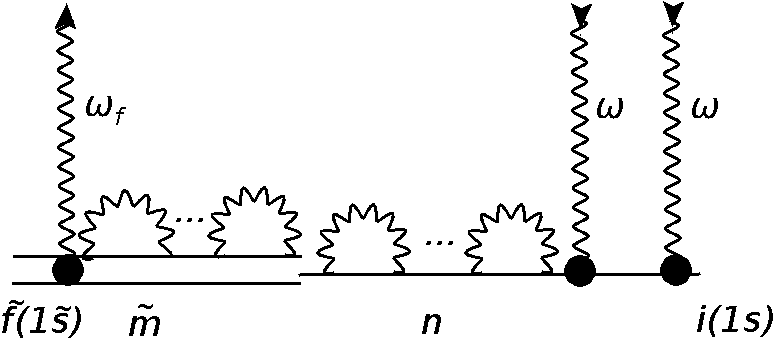}
\caption{
Schematic illustration of the subsequent insertions of the one-loop self-energy into the Feynman diagram Fig.~\ref{FigS1}. An infinite number of such insertions leads to a geometric progression that eventually gives the contribution to the energy denominator in the resonance terms of Eq. (\ref{s1.15}). The imaginary part of the one-loop self-energy correction represents the natural level width when averaged over unperturbed states $n$, and the level width stimulated by an external electric field (quadratic in the field) when averaged over perturbed states $\tilde{m}$. The real parts are represented by the Lamb shift and the quadratic Stark shift for particular states $n$ and $\tilde{m}$, respectively. The linear in-field contribution should go to zero after angular integration \cite{SSLP_2010,SS_2015} and therefore is not represented in the diagram.
 }
\end{figure}

Such inserts in the single solid line in Fig.~\ref{FigS2} are valid only for the $n=2p$ state. In this case, the natural width of the $2p$ level arises as the imaginary part of the self-energy operator \cite{LabKlim}. For the $2s$ state the picture is more complicated. In particular, it has recently been shown (see \cite{Jent3,jas08} and references therein) that the two-loop embedding can be used to substitute the two-photon level width into the resonance energy denominator. However, it should be emphasised that the "alternative approach" \cite{Jent3} is applicable for non-cascaded two-photon decays, see discussion in \cite{LSP_2009,ZSLP_2014}. In turn, the phenomenological approach of the level width insertions should be also considered accurately \cite{ZBSL_2014}.

Due to the smallness of the natural width of the $2s$ level, the insertions of the self-energy loops in the divergent energy denominator can only be regarded within the double solid line in Fig.~\ref{FigS2}. The loops overlapping the $\tilde{m}$ and $n$ states should lead to off-diagonal matrix element for the self-energy operator, the imaginary part of which is zero in the dipole approximation for the states $\tilde{m}={2\tilde{s}_{1/2},2\tilde{p}_{1/2}}$, $n= {2s_{1/2},2p_{1/2}}$. The diagonal self-energy matrix elements will correspond only to the higher photon multipoles, which we omit due to their smallness. Analytical calculations performed for the single insertion of the one-loop self-energy correction into $\tilde{m}$ line in Fig.~\ref{FigS2} for the resonant (first one for brevity) term in Eq. (\ref{s1.12}) can be performed as follows. 

The $S$-matrix element is
\begin{eqnarray}
\label{s1.16}
S_{fi}^{\rm SE} = (-\mathrm{i})^5 \int d^4x_1\dots d^4x_5 \overline{\psi}_f(x_1) \gamma^\mu A^*_\mu(x_1)
\tilde{S}(x_1,x_2)\frac{1-\boldsymbol{\alpha}_1\boldsymbol{\alpha}_2}{2\pi\mathrm{i}}\int\limits_{-\infty}^{\infty}d\kappa\frac{e^{\mathrm{i}|\kappa|r_{23}-\mathrm{i}\kappa(t_2-t_2)}}{r_{23}}\tilde{S}(x_2,x_3)
\\
\nonumber
\times 
S^{\rm FGS}(x_3,x_4)\gamma^{\nu_1}A_{\nu_1}(x_4)S(x_4,x_5)\gamma^{\nu_2}A_{\nu_2}\psi_i(x_5).
\end{eqnarray}
Here $\tilde{S}(x_1,x_2)$ denotes the Feynman electron propagator (\ref{s1.3}), but with decomposition by functions arising from the Hamiltonian with an external electric field (the out-Hamiltonian \cite{FGS91}). The photon loop operator (the Feynman photon propagator) is written out explicitly \cite{LabKlim}.

After integrating over the time variables using Eq. (\ref{s1.7}), and removing the integrals from the resulting $\delta$-functions, the expression (\ref{s1.16}) is simplified to
\begin{eqnarray}
\label{s1.17}
S_{fi} \approx - 2\pi\mathrm{i} e^3 \delta(E_f+\omega_f-2\omega-E_i) N
\sum\limits_{\tilde{n}_1,\tilde{m},n}\frac{A^{(1\gamma)}_{f\,\tilde{n}_1}}{E_f+\omega_f-E_{\tilde{n}_1}(1-\mathrm{i} 0)} 
\\
\nonumber
\times
\frac{1}{2\pi\mathrm{i}} 
\sum\limits_{\tilde{n}_2}\int\limits_{-\infty}^\infty d\kappa\frac{\langle \tilde{n}_1\tilde{n}_2| \frac{1-\vec{\alpha}_1\vec{\alpha}_2}{r_{12}}e^{\mathrm{i}|\kappa|r_{12}} |\tilde{n}_2\tilde{m}\rangle}{E_f+\omega_f-\kappa-E_{\tilde{n}_2}(1-\mathrm{i}0)}
\frac{\mathit{w}_{\tilde{m}\, n}  A^{(2\gamma)}_{n\,i}}{E_f+\omega_f-E_{\tilde{m}}(1-\mathrm{i}0)}.
\end{eqnarray}
By considering the diagonal contributions $\tilde{n}_1=\tilde{m}$, the operator highlighted by the $\times$ signs in the second line of Eq. (\ref{s1.17}) represent the one-loop self-energy contribution, which we hereafter denote as $\langle\tilde{m}|\hat{\Sigma}(E_f+\omega_f)|\tilde{m}\rangle$. The details of such an evaluation (as well as further accounting for the infinite number of loop insertions) can be found in \cite{Andr}.

Combining the expression (\ref{s1.17}) with the first (resonance) term in Eq. (\ref{s1.12}) for the particular state $\tilde{m}=2\tilde{s}$, we find 
\begin{eqnarray}
\label{s1.18}
S_{fi}^{\rm res} = -2\pi\mathrm{i} e^3 \delta(E_f+\omega_f-2\omega-E_i) N 
\frac{A^{(1\gamma)}_{f\,2\tilde{s}_{1/2}} \mathit{w}_{2\tilde{s}_{1/2}2s_{1/2}} A^{(2\gamma)}_{2s_{1/2}\, i}}{E_f+\omega_f-E_{2\tilde{s}}(1-\mathrm{i} 0)}\left[ 1 + \frac{\langle 2\tilde{s}|\hat{\Sigma}(E_f+\omega_f)|2\tilde{s}\rangle}{E_f+\omega_f-E_{2\tilde{s}}(1-\mathrm{i} 0)}\right].
\end{eqnarray}
Continuing in the same manner, one finds a series of geometric progression with a common ratio represented by the second summand in square brackets of Eq. (\ref{s1.18}). Then, summing the series and taking into account that $\langle\tilde{m}|\hat{\Sigma}(E_f+\omega_f)|\tilde{m}\rangle\approx \Delta E_{\tilde{m}}-\frac{\mathrm{i}}{2}\Gamma_{\tilde{m}}$, we arrive at the resonant amplitude
\begin{eqnarray}
\label{s1.19}
U_{fi}^{\rm res} =  N
\frac{A^{(1\gamma)}_{f\,2\tilde{s}_{1/2}} \mathit{w}_{2\tilde{s}_{1/2}2s_{1/2}} A^{(2\gamma)}_{2s_{1/2}\, i}}{E_f+\omega_f-E_{2\tilde{s}}-\Delta E_{2\tilde{s}}+\frac{\mathrm{i}}{2}\Gamma_{2\tilde{s}}}.
\end{eqnarray}
Here $\Delta E_{2\tilde{s}}$ is the quadratic Stark shift of the $2s$ state in an external electric field, and $\Gamma_{2\tilde{s}}$ is the field-stimulated level width. Estimates of the latter can be found in the \cite{Azimov1974,Mohr1978,SSLP_2010}, where the $2p$-state admixture is taken into account. 

The level width $\Gamma_{2\tilde{s}}$ can be presented in the form (see \cite{SSLP_2010} and references therein):
\begin{eqnarray}
\label{s1.20}
\Gamma_{2\tilde{s}} = W^{(1\gamma)}_{2\tilde{s}} + W^{(2\gamma)}_{2\tilde{s}}
\approx W^{(1\gamma)}_{2s} + W^{(2\gamma)}_{2s} + \frac{e^2\bm{E}^2 W^{(1\gamma)}_{2p}}{\Delta E_L^2+\frac{1}{4}\Gamma_{2p}^2} + \frac{e^2\bm{E}^2 W^{(2\gamma)}_{2p}}{\Delta E_L^2+\frac{1}{4}\Gamma_{2p}^2}.
\end{eqnarray}
The expression (\ref{s1.20}) consists of one- and two-photon contributions in the leading order of magnitude. It is well known that the one-photon decay of $2s$-states in the hydrogen atom is completely negligible and the natural width of these metastable states is formed by the two-photon $W^{(2\gamma)}_{ 2s}$ decay rate. However, the impurity of the $2p$ state gives rise to the third and fourth terms in Eq. (\ref{s1.20}), where the one- and two-photon decay rates of the $2p$ level are presented. Due to the smallness of the transition probability $W^{(2\gamma)}_{2p}$, it can be found that only the third term can be left in the expression for $\Gamma_{2\tilde{s}}$.

\section{Line profile}
\label{supp2}
To derive the line profile for the photon scattering process used in \cite{PhysRevLett.84.5496,Parthey,Mat}, we turn to the theory outlined in \cite{Jent-Mohr}. Taking the expression (\ref{s1.15}) as a starting point, fixing the final and initial states as $f=i=1s_{1/2}$, regularizing the divergent energy denominators according to Eq. (\ref{s1.19}), we also take into account $E_{2\tilde{p}}\approx E_{2p}\approx E_{2s}-\Delta E_L$ and $E_{2\tilde{ s}}\approx E_{2s}$. 
 Then the amplitude can be reduced to
\begin{eqnarray}
\label{s2.1}
U_{1s\,1s}\approx -e^3N\left[
 \frac{A^{(1\gamma)}_{1s\, 2p} \eta A^{(ext)}_{2p\, 2s}  A^{(2\gamma)}_{2s\,1s}}{x-\frac{\mathrm{i}}{2} \Gamma_{2\tilde{s}}}
+ \frac{A^{(1\gamma)}_{1s\, 2p}   A^{(2\gamma)}_{2p\,1s}}{x-\Delta E_L}  
+  \frac{A^{(1\gamma)}_{1s\, 2p} |\eta|^2 |A^{(ext)}_{2p\, 2s}|^2  A^{(2\gamma)}_{2p\,1s}}{x-\frac{\mathrm{i}}{2} \Gamma_{2\tilde{s}}}
- \frac{A^{(1\gamma)}_{1s\, 2p} \eta^* A^{(ext)}_{2p\, 2s}  A^{(2\gamma)}_{2s\,1s}}{x-\Delta E_L}
\right],\qquad
\end{eqnarray}
where $x\equiv E_{2s}-E_{1s}-2\omega = 2\omega_0-2\omega$, $A^{(ext)}_{2p\, 2s}\equiv \langle 2p| e\bm{E}\boldsymbol{r}|2s\rangle$ and the infinitesimal imaginary part in nonresonant energy denominators is discarded.

In order to obtain the cross section, the amplitude modulus (\ref{s2.1}) has to be squared, and hereafter the imaginary part of the factor $\eta=(\Delta E_L+\frac{\mathrm{i }}{2}\Gamma_{2p})^{-1}$ has to be taken into account. As a result, the resonance contribution is
\begin{eqnarray}
\label{s2.2}
\left|U_{1s\,1s}^{(\rm res)}\right|^2 = 
e^6N^2 |\eta|^4
\frac{\left|A_2 + A_1 \Delta E_L\right|^2 + \frac{\Gamma_{2p}^2\left|A_1\right|^2}{4} }{x^2+\frac{1}{4}\Gamma_{2\tilde{s}}},\qquad
\end{eqnarray}
where the notations $A_1=A^{(1\gamma)}_{1s\, 2p} \times A^{(ext)}_{2p\, 2s}  A^{(2\gamma)}_{2s\,1s}$ and $A_2=A^{(1\gamma)}_{1s\, 2p} \times |A^{(ext)}_{2p\, 2s}|^2  A^{(2\gamma)}_{2p\,1s}$ are introduced. In principle, the amplitude $A_2$ can be completely neglected since it consists of the square field strength multiplier and the amplitude of two-photon absorption into the $2p$ state, which is at least $\alpha$ (fine structure constant) times smaller than the amplitude of resonant absorption into the $2s$ state. Thus, 
\begin{eqnarray}
\label{s2.3}
\left|U_{1s\,1s}^{(\rm res)}\right|^2 \approx 
e^6N^2|\eta|^2
\frac{\left|A_1\right|^2 }{x^2+\frac{1}{4}\Gamma_{2\tilde{s}}}.\qquad
\end{eqnarray}

The following consideration corresponds to going beyond the resonance approximation. Then to the amplitude (\ref{s2.3}) one should add the squares of nonresonant contributions and terms interfering with the resonant one. The result can be presented in the form:
\begin{eqnarray}
\label{s2.4}
\left|U_{1s\,1s}^{(\rm nr)}\right|^2 (e^6N^2)^{-1}\approx \frac{\left|A_3\right|^2}{(\Delta E_L-x)^2} + \frac{|\eta|^2\left|A_4\right|^2}{(\Delta E_L-x)^2}
\\
\nonumber
+\left[\frac{\eta A_1}{x-\frac{\mathrm{i}}{2} \Gamma_{2\tilde{s}}}\left(\frac{A^*_3}{\Delta E_L-x}-\frac{\eta A^*_4}{\Delta E_L-x}\right)\right]
+\left[\frac{\eta^* A_1^*}{x+\frac{\mathrm{i}}{2} \Gamma_{2\tilde{s}}}\left(\frac{A_3}{\Delta E_L-x}-\frac{\eta^* A_4}{\Delta E_L-x}\right)\right],
\end{eqnarray}
where the symbol $*$ represents the complex conjugation, and $A_3=A^{(1\gamma)}_{1s\, 2p}   A^{(2\gamma)}_{2p\,1s}$, $A_4= A^{(1\gamma)}_{1s\, 2p} \times A^{(ext)}_{2p\, 2s}  A^{(2\gamma)}_{2s\,1s}$. We have kept the sum of the two summands for interference because the included amplitudes are complex.

To extract the real part of the expression in square brackets of Eq. (\ref{s2.4}) we employ the multipole decomposition \cite{SSLP_2010} of the photon wave function, Eq. (\ref{s1.2}). This procedure corresponds to replacing Dirac wave functions in the matrix element with Pauli ones and replacing the radiation operator by the following expressions:
\begin{eqnarray}
\label{s2.5}
(\boldsymbol{e}\boldsymbol{\alpha})e^{\mathrm{i}\boldsymbol{k}\boldsymbol{r}}\rightarrow \left(\boldsymbol{e}\boldsymbol{p}-\mathrm{i}(\boldsymbol{e}[\boldsymbol{k}\times\boldsymbol{s}])\right)e^{\mathrm{i}\boldsymbol{k}\boldsymbol{r}} \approx 
\mathrm{i}[\hat{H},(\boldsymbol{e}\boldsymbol{r})]-\frac{1}{2}[\hat{H},(\boldsymbol{e}\boldsymbol{r})(\boldsymbol{k}\boldsymbol{r})]-\frac{\mathrm{i}}{2}\left(\boldsymbol{e}[\boldsymbol{k}\times ([\boldsymbol{r}\times \boldsymbol{p}]+2\boldsymbol{s})]\right),
\\
\nonumber
(\boldsymbol{e}^*\boldsymbol{\alpha})e^{-\mathrm{i}\boldsymbol{k}\boldsymbol{r}}\rightarrow \left(\boldsymbol{e}^*\boldsymbol{p}+\mathrm{i}(\boldsymbol{e}^*[\boldsymbol{k}\times\boldsymbol{s}])\right)e^{-\mathrm{i}\boldsymbol{k}\boldsymbol{r}} \approx 
\mathrm{i}[\hat{H},(\boldsymbol{e}^*\boldsymbol{r})]+\frac{1}{2}[\hat{H},(\boldsymbol{e}^*\boldsymbol{r})(\boldsymbol{k}\boldsymbol{r})]+\frac{\mathrm{i}}{2}\left(\boldsymbol{e}^*[\boldsymbol{k}\times ([\boldsymbol{r}\times \boldsymbol{p}]+2\boldsymbol{s})]\right),
\end{eqnarray}
where the vector product is denoted by $\times$ in square brackets, the commutation relations with the Hamiltonian, $\hat{H}$, are given by square brackets with comma, and $\boldsymbol{p}$, $\boldsymbol{s}$ denote the momentum and spin momentum operators of the electron, respectively. The first summand in Eq. (\ref{s2.5}) represents the electric dipole of the photon, while the second and third represent the quadrupole and magnetic dipole, respectively. The vector product $[\boldsymbol{r}\times \boldsymbol{p}]=\boldsymbol{l}$ is the orbital momentum operator. The first line corresponds to the absorbed photon, whereas the second line represents the emission operator (the difference consists in the complex conjugation).

It can be found that the amplitude $A_4$ repeats $A_1$. Therefore, the products $A_1 A_4^*=A_1^*A_4 = |A_1|^2$ and
\begin{eqnarray}
\label{s2.6}
-\frac{\eta A_1}{x-\frac{\mathrm{i}}{2} \Gamma_{2\tilde{s}}}\frac{\eta^* A_4}{\Delta E_L-x} - \frac{\eta^* A_1^*}{x+\frac{\mathrm{i}}{2} \Gamma_{2\tilde{s}}}\frac{\eta^* A_4}{\Delta E_L-x} 
=  \frac{2 \left|A_1\right|^2 \left(x \Delta E_L^2-x\frac{\Gamma_{2p}^2}{4} + \Delta E_L \frac{\Gamma_{2p}\Gamma_{2\tilde{s}}}{2}\right)}{ 
(\Delta E_L-x) \left(x^2+\frac{1}{4}\Gamma_{2\tilde{s}}^2\right)}|\eta|^4
\end{eqnarray}

Considering the product $A_1A^*_3$, we can write
\begin{eqnarray}
\label{s2.7}
A_1A_3^* = A^{(1\gamma)}_{1s\, 2p} A^{(ext)}_{2p\, 2s}  A^{(2\gamma)}_{2s\,1s} \left(A^{(1\gamma)}_{1s\, 2p}   A^{(2\gamma)}_{2p\,1s}\right)^* 
= 
A^{(ext)}_{2p\, 2s} A^{(E1)}_{1s\, 2p}A^{(2E1)}_{2s\,1s} \times  A^{(E1)*}_{1s\, 2p} \left(A^{(E1M1)}_{2p\,1s} + A^{(E1E2)}_{2p\,1s}\right)^* =\qquad\qquad
\\
\nonumber
\langle 1s|[\hat{H},(\boldsymbol{e}^*_f\boldsymbol{r})]|2p\rangle A^{(ext)}_{2p\, 2s}\langle 2s| [\hat{H},(\boldsymbol{e}_1\boldsymbol{r})]|n_1\rangle\langle n_1|[\hat{H},(\boldsymbol{e}_2\boldsymbol{r})]|1s\rangle 
 \langle 1s| \frac{1}{2}\left(\boldsymbol{e}^*_2[(\boldsymbol{l}+2\boldsymbol{s})\times\boldsymbol{k}_2]\right)|n_2\rangle\langle n_2| [\hat{H},(\boldsymbol{e}^*_1\boldsymbol{r})]|2p\rangle \langle 2p| [\hat{H},(\boldsymbol{e}_f\boldsymbol{r})]| 1s\rangle
 \\
 \nonumber
+\mathrm{i}\langle 1s|[\hat{H},(\boldsymbol{e}^*_f\boldsymbol{r})]|2p\rangle A^{(ext)}_{2p\, 2s}\langle 2s| [\hat{H},(\boldsymbol{e}_1\boldsymbol{r})]|n_1\rangle\langle n_1|[\hat{H},(\boldsymbol{e}_2\boldsymbol{r})]|1s\rangle 
\langle 1s|\frac{1}{2}[\hat{H},(\boldsymbol{e}^*_2\boldsymbol{r})(\boldsymbol{k}_2\boldsymbol{r})]|n_2\rangle\langle n_2 | [\hat{H},(\boldsymbol{e}^*_1\boldsymbol{r})]|2p\rangle\langle 2p| [\hat{H},(\boldsymbol{e}_f\boldsymbol{r})]| 1s\rangle.
\end{eqnarray}
Here we should put $\boldsymbol{k}_{1,2}=\omega\boldsymbol{\nu}_{1,2}$ and consider that the matrix elements from the commutators $\langle a |[\hat {H},\hat{f}]|b\rangle = (E_a-~E_b)\langle a |\hat{f}|b\rangle$, which are converted to the corresponding frequencies $\omega_f$ or $\omega$ taking into account the photon permutations \cite{LabKlim}, and $A^{(ext)}_{2p\, 2s}$ represents the real multiplier. The result can be reduced to
\begin{eqnarray}
\label{s2.8}
A_1A_3^*=
\omega_f^2\omega^4\langle 1s|(\boldsymbol{e}^*_f\boldsymbol{r})|2p\rangle A^{(ext)}_{2p\, 2s}\langle 2s| (\boldsymbol{e}_1\boldsymbol{r})|n_1\rangle\langle n_1|(\boldsymbol{e}_2\boldsymbol{r})|1s\rangle
\langle 1s| \frac{1}{2}\left(\boldsymbol{e}^*_2[(\boldsymbol{l}+2\boldsymbol{s})\times\boldsymbol{\nu}_2]\right)|n_2\rangle\langle n_2| (\boldsymbol{e}^*_1\boldsymbol{r})|2p\rangle \langle 2p| (\boldsymbol{e}_f\boldsymbol{r})| 1s\rangle\qquad
\nonumber
\\
+\mathrm{i}\omega_f^2\omega^5\langle 1s|(\boldsymbol{e}^*_f\boldsymbol{r})|2p\rangle A^{(ext)}_{2p\, 2s}\langle 2s|(\boldsymbol{e}_1\boldsymbol{r})|n_1\rangle\langle n_1|(\boldsymbol{e}_2\boldsymbol{r})|1s\rangle 
\langle 1s|\frac{1}{2}(\boldsymbol{e}^*_2\boldsymbol{r})(\boldsymbol{\nu}_2\boldsymbol{r})|n_2\rangle\langle n_2 | (\boldsymbol{e}^*_1\boldsymbol{r})|2p\rangle\langle 2p| (\boldsymbol{e}_f\boldsymbol{r})| 1s\rangle.\qquad\qquad
\end{eqnarray}
Here and above, summation over $n_1,n_2$ is assumed, and terms with rearranged photons are discarded for brevity. In the same way, the complex conjugated product
\begin{eqnarray}
\label{s2.9}
A^*_1A_3=
\omega_f^2\omega^4\langle 1s|(\boldsymbol{e}^*_2\boldsymbol{r})|n_1\rangle\langle n_1|(\boldsymbol{e}^*_1\boldsymbol{r})|2s\rangle A^{(ext)}_{2s\, 2p}\langle 2p| (\boldsymbol{e}_f\boldsymbol{r})|1s\rangle
\langle 1s| (\boldsymbol{e}^*_f\boldsymbol{r})| 2p\rangle \langle 2p|(\boldsymbol{e}_1\boldsymbol{r}) |n_2\rangle\langle n_2| \frac{1}{2}\left(\boldsymbol{e}_2[(\boldsymbol{l}+2\boldsymbol{s})\times\boldsymbol{\nu}_2]\right)|1s\rangle \qquad
\\
\nonumber
-\mathrm{i}\omega_f^2\omega^5\langle 1s|(\boldsymbol{e}^*_2\boldsymbol{r})|n_1\rangle \langle n_1| (\boldsymbol{e}^*_1\boldsymbol{r})|2s\rangle A^{(ext)}_{2s\, 2p}\langle 2p|(\boldsymbol{e}_f\boldsymbol{r})|1s\rangle 
\langle 1s|(\boldsymbol{e}^*_f\boldsymbol{r})|2p\rangle\langle 2p | (\boldsymbol{e}_1\boldsymbol{r})|n_2\rangle\langle n_2| \frac{1}{2}(\boldsymbol{e}_2\boldsymbol{r})(\boldsymbol{\nu}_2\boldsymbol{r})| 1s\rangle,
\end{eqnarray}
i.e. is complex conjugate to $A_1A_3^*$.

The structure of the fourth amplitude shows that we can double the product $A_1^*A_4=|A_1|^2$ by separating the real part. Substituting Eqs. (\ref{s2.8}), (\ref{s2.9}) into the expression (\ref{s2.4}) the result can be written as
\begin{eqnarray}
\label{s2.10}
\left|U_{1s\,1s}^{(\rm nr)}\right|^2(e^6N^2)^{-1} \approx \frac{\left|A_3\right|^2}{(\Delta E_L-x)^2} + \frac{|\eta|^2\left|A_4\right|^2}{(\Delta E_L-x)^2}
+\frac{A_3^{(E1M1)} A_1^* \left(x \Delta E_L + \frac{1}{2}\Gamma_{2p} \Gamma_{2\tilde{s}}\right)}{
(\Delta E_L-x) \left(x^2+\frac{1}{4}\Gamma_{2\tilde{s}}^2\right)}|\eta|^2
\\
\nonumber
+\frac{A_3^{(E1E2)} A_1^* \left(\Delta E_L \Gamma_{2\tilde{s}} -x \Gamma_{2p}\right)}{(\Delta E_L-x) \left(x^2+\frac{1}{4}\Gamma_{2\tilde{s}}^2\right)}|\eta|^2
+ \frac{2 \left|A_1\right|^2 \left(x \Delta E_L^2-x\frac{\Gamma_{2p}^2}{4} + \Delta E_L \frac{\Gamma_{2p}\Gamma_{2\tilde{s}}}{2}\right)}{ 
(\Delta E_L-x) \left(x^2+\frac{1}{4}\Gamma_{2\tilde{s}}^2\right)}|\eta|^4,
\end{eqnarray}
where $A_3^{(E1M1)}$ denotes the part corresponding to the E1M1 two-photon transition.

Then using the decomposition over the small parameter $x/\Delta E_L$ and dropping out the $x$ independent terms, in the linear over $x$ approximation the Fano profile can be obtained as
\begin{eqnarray}
\label{s2.11}
\left|U_{1s\,1s}\right|^2\approx  \frac{C}{x^2+\frac{1}{4}\Gamma_{2\tilde{s}}}
+ a\, x
+\frac{b\, x}{x^2+\frac{1}{4}\Gamma_{2\tilde{s}}^2},
\end{eqnarray}
with the notations introduced as in \cite{Jent-Mohr}. In our case
\begin{eqnarray}
\label{s2.12}
C=e^6N^2|\eta|^2\left(\left|A_1\right|^2  + A_3^{(E1M1)} A_1^* \frac{\Gamma_{2p} \Gamma_{2\tilde{s}} }{2\Delta E_L} + A_3^{(E1E2)} A_1^*\Gamma_{2\tilde{s}}+ \left|A_1\right|^2  \Gamma_{2p} \Gamma_{2\tilde{s}} |\eta|^2\right),
\nonumber
\\
a= e^6N^2\left(\frac{\left|A_3\right|^2 }{\Delta E_L^3}+\frac{\left|A_1\right|^2 |\eta|^2 }{\Delta E_L^3}\right),\qquad\qquad
\\
\nonumber
b =  e^6N^2\left( A_3^{(E1M1)} A_1^* |\eta|^2 \left(1+\frac{\Gamma_{2p} \Gamma_{2\tilde{s}}}{2\Delta E_L^2}\right)  
+  A_3^{(E1E2)} A_1^* |\eta|^2 \frac{\Gamma_{2\tilde{s}}-\Gamma_{2p}}{\Delta E_L} +
2 \left|A_1\right|^2 \frac{\left|\eta\right|^4}{\Delta E_L}\left(\Delta E_L^2-\frac{\Gamma_{2p}^2}{4}
 + \frac{\Gamma_{2p} \Gamma_{2\tilde{s}}}{2}\right)\right).
\end{eqnarray}
To check the dimensionality of expressions in (\ref{s2.12}), it is necessary to remember that the amplitude $A_1$ contains an additional energy multiplier $A_{2p\, 2s}^{(ext)}=\langle 2p|e \bm{E}\boldsymbol{r}|2s\rangle$.

The line profile (\ref{s2.11}) can be rewritten (in linear approximation) in the following form:
\begin{eqnarray}
\label{s2.13}
\left|U_{1s\,1s}\right|^2\approx \frac{C}{\left[x-\Delta(x)\right]^2+\frac{1}{4}\Gamma_{2\tilde{s}}^2},\qquad
\\
\nonumber
\Delta(x) = \frac{a}{2C}\left(x^2+\frac{1}{4}\Gamma_{2\tilde{s}}^2\right)^2 + \frac{b}{2C}\left(x^2+\frac{1}{4}\Gamma_{2\tilde{s}}^2\right).
\end{eqnarray}
Thus, the asymmetric Fano contour (\ref{s2.11}) reduced to the expression (\ref{s2.13}) can be used to determine the energy shift at a given value of frequency. At the maximum of the line $x=0$ and at the full-width half-maximum $x=\Gamma_{2\tilde{s}}/2$, the frequency shift is equal to
\begin{eqnarray}
\label{s2.14}
\Delta(0) = \frac{b \Gamma_{2\tilde{s}}^2}{8 C} + \frac{a \Gamma_{2\tilde{s}}^4}{32 C},
\\
\nonumber
\Delta\left(\pm\frac{\Gamma_{2\tilde{s}}}{2}\right) = \frac{b \Gamma_{2\tilde{s}}^2}{4 c} + \frac{a \Gamma_{2\tilde{s}}^4}{16 C},
\end{eqnarray}
the smallest of them corresponds to the maximum of the line shape. The coefficients $a$, $b$, $C$ can be calculated using the theory provided in the following Appendix~\ref{supp3} and the theory for calculating the level width $\Gamma_{2\tilde{s}}$ presented in \cite{SSLP_2010,SS_2015}.

It should be noted that the expression (\ref{s2.14}) obtained for the frequency shift does not lead to the estimates given earlier \cite{LShSP_2007,LShSP_PRL_2007,LSSCK_2009} for two reasons. The first one refers to the presence of an imaginary part in the mixing coefficient of states $2s$ and $2p$, see Eqs. (\ref{s1.9}), (\ref{s1.10}). The second is the consideration of complexity in the multipole decomposition of the photon wave function. For completeness, we can also point out that the first summand in (\ref{s2.12}) for $C$ can be replaced according to expression (\ref{s2.2}), i.e. $\left|A_1\right|^2\rightarrow |\eta|^2\left(|A_2+A_1\Delta E_L|^2+\left|A_1\right|^2\Gamma_{2p}^2/4\right)$.

\section{Angular integration of the scattering amplitudes}
\label{supp3}

The Fano profile of Eq. (\ref{s2.11}) allows the line shape asymmetry to be picked out. Recently, in \cite{H-exp} it was this contour that greatly improved the accuracy of the transition frequency determination, see also \cite{udem2019_AND}. For this purpose, the parameters $a,b$ and $C$ were used to fit the experimental data. It should be noted that even for the absorption frequency $2s-4p$, the asymmetry is not observable to the naked eye, but has a significant effect on the value of the transition frequency (the asymmetry is significant due to the presence of a close state of the same parity as the resonance one and also cascade emission processes \cite{SAZL_PRA2024,SAZL_PhysRevA2024}). A considerable effect due to the asymmetry of the line profile can hardly be expected for the frequency $2s-1s$, but at least it can be of particular importance in order to further improve the accuracy.

To estimate the asymmetry of the observed line profile in the experiments \cite{PhysRevLett.84.5496,Parthey,Mat}, we provide below an analytical calculation of the quantities included in the coefficients $a,b$ and $C$ of Eq. (\ref{s2.11}). We first obtain the square of the amplitude modulus for the first (resonance) term in Eq. (\ref{s2.1}). According to the notations introduced in the previous section, we have
\begin{eqnarray}
\label{s3.1}
\left|A_1\right|^2 = A^{(1\gamma)}_{1s\, 2p} A^{(ext)}_{2p\, 2s}  A^{(2\gamma)}_{2s\,1s}
\times 
A^{(2\gamma)*}_{1s\,2s}A^{(ext)*}_{2s\, 2p}A^{(1\gamma)*}_{2p\, 1s}.\qquad
\end{eqnarray}

To perform angular algebra in Eq. (\ref{s3.1}) one can use the Wigner-Eckart theorem, see, e.g., \cite{Varsh}. To accomplish this, we take into account the hyperfine structure of the states. Then the states are given by the set of quantum numbers $nljFM$, where $n$ is the principal quantum number, $l$ is the orbital momentum of the electron, $j$ is the total angular momentum summed from the orbital and spin, $s$, momenta, and $F$ is the total atomic momentum accounting for the spin of the nucleus, $I$ ($I=1/2$ for the hydrogen atom as well as the electron spin). Using such a set of quantum numbers, it is possible to determine the amplitude Eq.~(\ref{s3.1})
\begin{eqnarray}
    \label{ss3.1}
    \langle 1s^{F_f}_{1/2} | \boldsymbol{e}^*_f \boldsymbol{r} | 2p^{F_p}_{1/2} \rangle \langle 2p^{F_p}_{1/2} | \boldsymbol{Er} | 2s^{F_s}_{1/2} \rangle \langle 2s^{F_s}_{1/2} | \boldsymbol{e}_c \boldsymbol{r} | k^{F_k}_{j_k} \rangle 
\langle k^{F_k}_{j_k} | \boldsymbol{e}_{\bar{c}}\boldsymbol{r} | 1s^{F_i}_{1/2} \rangle \langle 1s^{F_i}_{1/2} | \boldsymbol{e}^*_{t}\boldsymbol{r} | n^{F_n}_{j_n} \rangle
    \\
    \nonumber
    \times 
\langle n^{F_n}_{j_n} | \boldsymbol{e}^*_{\bar{t}}\boldsymbol{r} | 2s^{F_{s^{\prime}}}_{1/2} \rangle
\langle 2s^{F_{s^{\prime}}}_{1/2} | \boldsymbol{Er} | 2p^{F_{p^{\prime}}}_{1/2} \rangle \langle 2p^{F_{p^{\prime}}}_{1/2} | \boldsymbol{e}^*_f \boldsymbol{r} | 1s^{F_i}_{1/2} \rangle.
\end{eqnarray}
In the expression above indices $c = 1,2$ and $\bar{c} = 1,2$ while $c \neq \bar{c}$, i.e. $\bar{c}$ is the 'opposite' to $c$, the same holds for $t$ and $\bar{t}$. Combinations of these indices will correspond to photons permutations. The states denoted using $n$ and $k$ related to the spectra summations from the electron propagator.

It is convenient to perform angular integration using cyclic coordinates, for which the scalar product of two arbitrary vectors can be written as
\begin{eqnarray}
    \label{ss3.2}
    (\boldsymbol{a} \cdot \boldsymbol{b} ) \equiv \boldsymbol{ab} = \sum_q (-1)^q a_{q} b_{-q},
\end{eqnarray}
where $a_q$ and $b_q$ are cyclic components. We also use the following definition of irreducible tensor product \cite{Varsh}
\begin{eqnarray}
    \label{ss3.3}
    \{ \boldsymbol{a} \otimes \boldsymbol{b} \}_{x\xi} = \sum_{q_1q_2} C^{x\xi}_{1q_1 1q_2} a_{q_1} b_{q_2},
\end{eqnarray}
or, in case of the two arbitrary irreducible tensors product
\begin{eqnarray}
    \label{ss3.4}
    \{ \boldsymbol{A}_a \otimes \boldsymbol{B}_b \}_{c\gamma} = \sum_{\alpha \beta} C^{c\gamma}_{a \alpha b \beta} A_{a \alpha} B_{b \beta}.
\end{eqnarray}
The complex conjugation of the tensor product of two arbitrary vectors is given by
\begin{eqnarray}
    \label{ss3.5}
    \{ \boldsymbol{a} \otimes \boldsymbol{b} \}^*_{x\xi} = (-1)^{x - \xi} \{ \boldsymbol{a}^* \otimes \boldsymbol{b}^* \}_{x-\xi}.
\end{eqnarray}

Matrix element of the cyclic component $r_q$ of the radius vector has the following form:
\begin{eqnarray}
\label{ss3.6}
\langle n'l'j'F'M'|r_{q}|nljFM\rangle = 
(-1)^{F'-M'}
\begin{pmatrix}
     F' & 1 & F \\
-M' & q & M
\end{pmatrix}
\langle n'l'j'F'||r||nljF\rangle
\end{eqnarray}
where the Wigner $3jm$-symbol is introduced, and the reduced matrix element is
\begin{eqnarray}
\label{ss3.7}
\langle n'l'j'F'||r||nljF\rangle= (-1)^{j'+j+I+l'+1/2+F}
\Pi_{j'jF'F}
\begin{Bmatrix}
j' & F' & I \\
F  & j  & 1
\end{Bmatrix}
\begin{Bmatrix}
l' & j' & 1/2 \\
j  & l  & 1
\end{Bmatrix}
\langle n' l' || r|| nl \rangle,\qquad
\end{eqnarray}
and
\begin{eqnarray}
\label{ss3.8}
\langle n' l' || r || nl \rangle = (-1)^{l'}\Pi_{l'l}
\begin{pmatrix}
l' & 1 & l\\
0 & 0 & 0
\end{pmatrix}
\int_{0}^{\infty}dr\,r^3 R_{n'l'}R_{nl}.\qquad
\end{eqnarray}
In Eq. (\ref{ss3.8}) $ R_{nl} $ denotes the radial part of hydrogen wave function, $\Pi_{ab\dots}=\sqrt{(2a+1)(2b+1)\dots}$ and $\begin{Bmatrix}
j_1 & j_2 & j_3 \\
j_4  & j_5  & j_6
\end{Bmatrix}$ is the Wigner $6j$-symbol.

Using the above definitions and formulae one obtains for $|A_1|^2$
\begin{eqnarray}
    \label{ss3.9}
    (-1)^{\sum q + \sum q^{\prime}} (e^*_f)_{-q_1} (E)_{-q_2} (e_c)_{-q_3} (e_{\bar{c}})_{-q_4} (e^*_t)_{-q^{\prime}_4} (e^*_{\bar{t}})_{-q^{\prime}_3} (E)_{-q^{\prime}_2} (e_f)_{-q^{\prime}_1}
    \\
    \nonumber
    \times
    (-1)^{F_f - M_f}
    \begin{pmatrix}
        F_f & 1 & F_p \\
        -M_f & q_1 & M_p
    \end{pmatrix}
    \langle 1s || r || 2p_{1/2} \rangle
    (-1)^{F_p - M_p}
    \begin{pmatrix}
        F_p & 1 & F_s \\
        -M_p & q_2 & M_s
    \end{pmatrix}
    \langle 2p_{1/2} || r || 2s \rangle
    \\
    \nonumber
    \times
    (-1)^{F_s - M_s}
    \begin{pmatrix}
        F_s & 1 & F_k \\
        -M_s & q_3 & M_k
    \end{pmatrix}
    \langle 2s || r || k \rangle
    (-1)^{F_k - M_k}
    \begin{pmatrix}
        F_k & 1 & F_i \\
        -M_k & q_4 & M_i
    \end{pmatrix}
    \langle k || r || 1s \rangle
    \\
    \nonumber
    \times
    (-1)^{F_i - M_i}
    \begin{pmatrix}
        F_i & 1 & F_n \\
        -M_i & q^{\prime}_4 & M_n
    \end{pmatrix}
    \langle 1s || r || n \rangle
    (-1)^{F_n - M_n}
    \begin{pmatrix}
        F_n & 1 & F_{s^{\prime}} \\
        -M_n & q^{\prime}_2 & M_{s^{\prime}}
    \end{pmatrix}
    \langle n || r || 2s \rangle
    \\
    \nonumber
    \times
    (-1)^{F_{s^{\prime}} - M_{s^{\prime}}}
    \begin{pmatrix}
        F_{s^{\prime}} & 1 & F_{p^{\prime}} \\
        -M_{s^{\prime}} & q^{\prime}_2 & M_{p^{\prime}}
    \end{pmatrix}
    \langle 2s || r || 2p_{1/2} \rangle
    (-1)^{F_{p^{\prime}} - M_{p^{\prime}}}
    \begin{pmatrix}
        F_{p^{\prime}} & 1 & F_i \\
        -M_{p^{\prime}} & q^{\prime}_1 & M_i
    \end{pmatrix}
    \langle 2p_{1/2} || r || 1s \rangle.
\end{eqnarray}
In the expression above the phase $\sum q + \sum q^{\prime}$ corresponds to all scalar products and contains vector components indices from the first line. By employing the relation \cite{Varsh}
\begin{eqnarray}
\label{ss3.10}
\sum\limits_\kappa (-1)^{q-\kappa}
\begin{pmatrix}
   a & b & q \\
   \alpha & \beta & -\kappa
\end{pmatrix}
\begin{pmatrix}
   q & d & c \\
   \kappa & \delta & \gamma
\end{pmatrix} 
 = (-1)^{2a}\sum\limits_{x\xi}
 (-1)^{x-\xi}\Pi^2_x
\begin{pmatrix}
   a & c & x \\
   \alpha & \gamma & -\xi
\end{pmatrix}
\begin{pmatrix}
   x & d & b \\
   \xi & \delta & \beta
\end{pmatrix}
\begin{Bmatrix}
b & d & x \\
c  & a  & q
\end{Bmatrix},
\end{eqnarray}
one can rearrange the indices in the $3jm$-symbols so that the cyclic components of the vectors in the first line of Eq. (\ref{ss3.9}) are converted to the tensor products in a convenient for further evaluations form. By summing Eq.~(\ref{ss3.9}) over all total angular momenta projections and utilizing Eq.~(\ref{ss3.10}) one obtains a final form of the tensor product that represents angular dependence of the line profile. We also perform summation of the $6j$-symbols over the following angular momenta quantum numbers: $F_p, F_{p^{\prime}},F_k,F_n$ and $j_k,j_n$. Since in this paper we consider only the mixing of $2s_{1/2}$ and $2p_{1/2}$ states, the quantum numbers $j_p = 1/2$ and $j_{p^{\prime}} = 1/2$ are fixed. The same is true for $F_i = 1$ and $F_s = 1$, $F_{s^{\prime}} = 1$, since the experiments \cite{PhysRevLett.84.5496,Parthey,Mat} study the transition between specific hyperfine components $1s^{F = 1}_{1/2} \rightarrow 2s^{F = 1}_{1/2}$. 

For brevity, we omit here intermediate calculations and present only the final result
\begin{eqnarray}
    \label{3.11}
    \sum_{F_f} (-1)^{\varphi} \Pi^2_{F_f F_i F_s F_{s^{\prime}}} \Pi^2_{j_f j_i j_s j_{s^{\prime}} j_p j_{p^{\prime}}} \langle 1s || r || 2p \rangle ^2 \langle 2p || r || 2s \rangle ^2 \langle 2s || r || k \rangle \langle k || r || 1s \rangle \langle 1s || r || n \rangle \langle n || r || 2s \rangle\qquad
    \\
    \nonumber
    \times
    \begin{Bmatrix}
        x & z & y \\
        F_i  & F_s  & F_f
    \end{Bmatrix}
    \begin{Bmatrix}
        x^{\prime} & z & y^{\prime} \\
        F_i  & F_{s^{\prime}}  & F_f
    \end{Bmatrix}
    \begin{Bmatrix}
        j_f & x & j_s \\
        F_s  & 1/2  & F_f
    \end{Bmatrix}
    \begin{Bmatrix}
        j_f & x^{\prime} & j_{s^{\prime}} \\
        F_{s^{\prime}}  & 1/2  & F_f
    \end{Bmatrix}
    \begin{Bmatrix}
        y & j_s & j_i \\
        1/2  & F_i  & F_s
    \end{Bmatrix}
    \begin{Bmatrix}
        y^{\prime} & j_i & j_{s^{\prime}} \\
        1/2  & F_{s^{\prime}}  & F_i
    \end{Bmatrix}
    \begin{Bmatrix}
        l_s & y & l_i \\
        1  & l_k  & 1
    \end{Bmatrix}
    \begin{Bmatrix}
        l_s & y & l_i \\
        j_i  & 1/2  & j_s
    \end{Bmatrix}
    \\
    \nonumber
    \times
    \begin{Bmatrix}
        l_i & y^{\prime} & l_{s^{\prime}} \\
        1  & l_n  & 1
    \end{Bmatrix}
    \begin{Bmatrix}
        l_i & y^{\prime} & l_{s^{\prime}} \\
        j_{s^{\prime}}  & 1/2  & j_i
    \end{Bmatrix}
    \begin{Bmatrix}
        1 & j_f & j_p \\
        1/2  & l_p  & l_f
    \end{Bmatrix}
    \begin{Bmatrix}
        1/2 & l_p & j_p \\
        1  & j_s  & l_s
    \end{Bmatrix}
    \begin{Bmatrix}
        1 & j_s & j_p \\
        j_f  & 1  & x
    \end{Bmatrix}
    \begin{Bmatrix}
        1 & j_f & j_{p^{\prime}} \\
        1/2  & l_{p^{\prime}}  & l_f
    \end{Bmatrix}
    \begin{Bmatrix}
        1/2 & l_{p^{\prime}} & j_{p^{\prime}} \\
        1  & j_{s^{\prime}}  & l_{s^{\prime}}
    \end{Bmatrix}
    \begin{Bmatrix}
        1 & j_{s^{\prime}} & j_{p^{\prime}} \\
        j_f  & 1  & x^{\prime}
    \end{Bmatrix}
    \\
    \nonumber
    \times
    (-1)^{x + x^{\prime}} \Pi_{xy x^{\prime} y^{\prime}} 
    \{ \{ \bm{E} \otimes \boldsymbol{e}^*_f \}_x \otimes \{ \boldsymbol{e}_{\bar{c}} \otimes \boldsymbol{e}_c \}_y \}_z \cdot \{ \{ \bm{E} \otimes \boldsymbol{e}^*_f \}_{x^{\prime}} \otimes \{ \boldsymbol{e}_{t} \otimes \boldsymbol{e}_{\bar{t}} \}_{y^{\prime}} \}_z
\end{eqnarray}
In the above expression reduced matrix elements are defined by Eq.~(\ref{ss3.8}). Phase $\varphi = F_s + F_{s^{\prime}} + l_i + l_f + j_f + l_p + 1 + j_s + l_s + j_f + j_{s^{\prime}} + l_{s^{\prime}}$ and summation over $x,y,x^{\prime},y^{\prime}$ and $z$ is assumed. Summation over $kl_k$ and $nl_n$ from electron propagator is also presumed in this expression, though according to the selection rules only $l_k = l_n = 0$ are left. We also note that the tensors that form the scalar product can differ only in the order of the photon indices and introduce the notation
\begin{eqnarray}
    U_{xyz} = \{ \{ \bm{E} \otimes \boldsymbol{e}^*_f \}_x \otimes \{ \boldsymbol{e}_{\bar{c}} \otimes \boldsymbol{e}_c \}_y \}_z.
\end{eqnarray}

By setting all quantum numbers in accordance to the transition studied in \cite{PhysRevLett.84.5496,Parthey,Mat}, one obtains the following factor for the $|A_1|^2$
\begin{eqnarray}
    \label{ss3.12}
    \frac{1}{81}\left( 3 U_{000} \cdot U_{000} + 2 U_{101} \cdot U_{101} \right).
\end{eqnarray}
Using the formulae from \cite{Varsh} one finds 
\begin{eqnarray}
    \label{ss3.13}
    U_{000} = \frac{1}{3} (\bm{E} \boldsymbol{e}^*_f) ( \boldsymbol{e}_c \boldsymbol{e}_{\bar{c}} ),
    \\
    \nonumber
    U_{101} = -\frac{i}{\sqrt{6}} [\bm{E} \times \boldsymbol{e}^*_f] ( \boldsymbol{e}_c \boldsymbol{e}_{\bar{c}} ).
\end{eqnarray}
At this point we consider two situations. The first one corresponds to the case, when photon polarizations are fixed. We assume that in experiment there are no effects that change the photons' polarization directions and polarizations are linear ($\boldsymbol{e}^* = \boldsymbol{e}$). Since the two absorbed photons come from the same laser $\boldsymbol{e}_1 = \boldsymbol{e}_2$. By denoting the angle between $\bm{E}$ and $\boldsymbol{e}^*_f$ as $\theta$ Eq.~(\ref{ss3.12}) and (\ref{ss3.13}) result in
\begin{eqnarray}
    \label{ss3.14}
    U_{000} = \frac{1}{3} D \cos \theta,
    \\
    \nonumber
    U_{101} = -\frac{i}{\sqrt{6}}  [\bm{E} \times \boldsymbol{e}^*_f] \sim \sin \theta,
    \\
    \nonumber
    \frac{1}{81}\left( 3 U_{000} \cdot U_{000} + 2 U_{101} \cdot U_{101} \right) = \frac{1}{243} D^2,
\end{eqnarray}
where $D = |\bm{E}|$ is the external electric field strength. The last expression in Eq.~(\ref{ss3.14}) means that the amplitude $|A_1|^2$ contains no angular correlations and proportional to the $D^2$.

Exactly similar result holds for the second situation, when polarization of the absorbed photons is not fixed, so that contribution of all polarization directions should be taken into account. This can be done by utilizing the following expression (in Cartesian components)
\begin{eqnarray}
\label{ss3.15}
\sum\limits_{\boldsymbol{e}}e_ie^*_k = \delta_{ik}-\nu_i\nu_k.
\end{eqnarray}

To accurately account for the asymmetry of the line contour one should also consider the amplitude $A_1 A^*_3$ representing interference with the two-photon absorption of E1M1 and E1E2 photons, see Eqs. (\ref{s2.10}), (\ref{s2.12}). In both parts of this expression one has terms that are linear in photon (either $1$ or $2$) propagation vector. According to the experimental scheme \cite{PhysRevLett.84.5496,Parthey,Mat} absorbed photons have opposite directions $\boldsymbol{\nu}_1 = -\boldsymbol{\nu}_2$. It could be shown that for both cases above (fixed and arbitrary polarizations) when all permutations of the absorbed photons is taken into account, the contribution of $A_1 A^*_3$ vanishes. 

In case when polarization vectors of the two absorbed photons are fixed, expression $A_1 A^*_3$ is linear in $\boldsymbol{\nu}_{1,2}$. Under the assumption $\boldsymbol{e}_1 = \boldsymbol{e}_2$, permutations of photons do not change $A_1$. Since permutation $1 \leftrightarrow 2$ leads to the emergence of a minus sign in $A_3$ due to $\boldsymbol{\nu}_1 = -\boldsymbol{\nu}_2$, the result for all permutations will be zero. In the case when summation over polarizations should be performed one can use the cartesian coordinates to obtain the following
\begin{eqnarray}
    \label{eeee}
    \sum_{\boldsymbol{e}_1 \boldsymbol{e}_2} \boldsymbol{e}_{1,i}\boldsymbol{e}_{2,k}\boldsymbol{e}^{*}_{1,s}\boldsymbol{e}^{*}_{2,r} = \delta_{is}\delta_{kr} + \nu_{1,i} \nu_{1,s} \nu_{2,k} \nu_{2,r} -
    \delta_{is} \nu_{2,k} \nu_{2,r} - \delta_{kr} \nu_{1,i} \nu_{1,s},
\end{eqnarray}
In this expression permutation $1 \leftrightarrow 2$ is always 'even', i.e. does not change the sign of the expression. In the same time, amplitude $A_1 A^*_3$ still contains $\boldsymbol{\nu}_{1,2}$, which makes it 'odd' under the permutations, giving the zero results for $A_1 A^*_3$. We additionally note that since contribution of $A_1 A^*_3$ vanishes, expression for the frequency shift will contain only $|A_1|^2$ in $b$ and $C$ coefficients from Eq.~(\ref{s2.12}), that will cancel itself in the nominator and denominator. 

The resulting asymmetry shift can be expressed as
\begin{eqnarray}
\label{ss3.24}
\frac{b}{2C} = \frac{4\Delta E_L^2-\Gamma_{2p}(\Gamma_{2p}-2\Gamma_{2\tilde{s}})}{4\Delta E_L^3+\Delta E_L\Gamma_{2p}(\Gamma_{2p}+4\Gamma_{2\tilde{s}})}.
\end{eqnarray}
For the $\Delta(x)$ defined at maximum and full-width half-maximum of the line profile, it leads to
\begin{eqnarray}
\label{ss3.25}
\Delta(0) = \frac{b}{2C}\frac{\Gamma_{2\tilde{s}}^2}{4} = [0.46;~ 7.4]\text{ Hz},
\\
\nonumber
\Delta\left(\pm \frac{\Gamma_{2\tilde{s}}}{2}\right) = \frac{b}{2C}\frac{\Gamma_{2\tilde{s}}^2}{2} = [0.9;~ 14.8]\text{ Hz},
\end{eqnarray}
where values are given at the field strength $10$ and $20$ V/cm in conjunction with estimation $\Gamma_{2\tilde{s}}\approx \left(\frac{\bm{E}}{475 \text{V/cm}}\right)^2\Gamma_{2p}$.

\section{Non-adiabatic electric field switching and modified line profile}
\label{supp5}

The wave functions in Eq.~(\ref{s1.9}) are obtained from perturbation theory \cite{Azimov1974,Mohr1978} for the case of a constant external electric field defined throughout all space. We now consider the scenario where the perturbation is switched on in a region separated from the excitation zone, with a time delay $t = \tau$:
\begin{eqnarray}
	\label{s5.1}
	V(t) = 
	\begin{pmatrix}
		0 & A^{(ext)}_{2p,2s} \\
		A^{(ext)}_{2s,2p} & 0
	\end{pmatrix}
	\theta(t - \tau).
\end{eqnarray}
In this expression, we take into account that $\left( A^{(ext)}_{2p,2s}\right)^* = A^{(ext)}_{2s,2p}$, where $A^{(ext)}_{2p,2s}\equiv \langle 2p| e\bm{E}\cdot\bm{r}|2s\rangle$. The matrix in Eq.~(\ref{s5.1}) is written in the basis of the zero-field Hamiltonian eigenstates $|2s\rangle$ and $|2p\rangle$. To find the wave function of the mixed $2\tilde{s}_{1/2}$ (or $2\tilde{p}_{1/2}$) state, we employ time-dependent perturbation theory (see, e.g., \cite{Landau}):
\begin{eqnarray}
	\label{s5.2}
	| 2\tilde{s}_{1/2} \rangle = c_{2s} (t) | 2s \rangle + c_{2p} (t) | 2p \rangle.
\end{eqnarray}
Time-dependent coefficients are found by solving the system
\begin{eqnarray}
	\label{s5.3}
	\begin{cases}
		i \dot{c}_{2s} (t) = A^{(ext)}_{2p,2s} \theta(t - \tau) e^{i \omega_{2s2p} t} c_{2p}(t),\\
		i \dot{c}_{2p} (t) = A^{(ext)}_{2s,2p} \theta(t - \tau) e^{-i \omega_{2s2p} t} c_{2s}(t).
	\end{cases}
\end{eqnarray}
In the preceding equations, $\omega_{2s2p} \equiv E_{2s} - E_{2p} = \Delta E_L$. We impose initial conditions such that before the electric field is turned on, the electron is in the $2s$ state: $c_{2s}(\tau^-) = 1$, $c_{2p}(\tau^-) = 0$. The solution of Eq.~(\ref{s5.3}) for $t < \tau$ is straightforward, while for $t \geq \tau$ it yields
\begin{eqnarray}
	\label{s5.4}
	c_{2s}(t) = e^{i(t - \tau) \Delta E_L / 2} \left\{ \cos \left[ \frac{1}{2}\sqrt{1 + 4 \epsilon^2} (t - \tau) \Delta E_L \right] - \frac{i}{\sqrt{1 + 4 \epsilon^2}} \sin \left[ \frac{1}{2}\sqrt{1 + 4 \epsilon^2} (t - \tau) \Delta E_L \right]  \right\},
	\nonumber
	\\
	c_{2p}(t) = 2 i e^{- i (t - \tau) \Delta E_L / 2}\frac{ \epsilon}{\sqrt{1 + 4 \epsilon^2}} \sin \left[ \frac{1}{2}\sqrt{1 + 4 \epsilon^2} (t - \tau) \Delta E_L \right],
\end{eqnarray}
where we have introduced parameter $\epsilon = |A^{(ext)}_{2p,2s}| / \Delta E_L$. As a result, the coefficients from Eq.~(\ref{s1.10}) now take a different form, incorporating $c_{2s}(t)$ and $c_{2p}(t)$. Specifically, $w_{2\tilde{s}_{1/2}2s_{1/2}} = c_{2s}(t)$ and $w_{2\tilde{s}_{1/2}2p_{1/2}} = c_{2p}(t)$.

In Appendix~\ref{supp3}, it was shown that when all photon permutations are taken into account, the numerator $C$ of the line profile in Eq.~(\ref{s2.13}) contains only $|A_1|^2$, corresponding to the first term in Eq.~(\ref{s1.12}). Recall that $A_1 = A^{(1\gamma)}_{f,2\tilde{s}_{1/2}} w_{2\tilde{s}_{1/2}2s_{1/2}} A^{(2\gamma)}_{2s_{1/2}, i}$. With the new mixing coefficients, this amplitude transforms to $A_1 = c_{2p}(t) A^{(1\gamma)}_{1s_{1/2},2p_{1/2}} c_{2s}(t) A^{(2\gamma)}_{2s_{1/2}, 1s_{1/2}}$, where we have set $f = i = 1s_{1/2}$. The squares of the absolute values of $c_{2s}(t)$ and $c_{2p}(t)$ yield
\begin{eqnarray}
	\label{s5.6}
	|c_{2s}(t)|^2 = \frac{1}{1 + 4 \epsilon^2} \left( 1 + 2\epsilon^2 + 2\epsilon^2 \cos \left[ \sqrt{1 + 4 \epsilon^2} (t - \tau) \Delta E_L \right] \right) =1 + 2\left( 1 - \cos \left[ (t - \tau) \Delta E_L \right] \right) \epsilon^2 + \mathcal{O}(\epsilon^4),
	\\
	\nonumber
	|c_{2p}(t)|^2 = \frac{2 \epsilon^2}{1 + 4 \epsilon^2} \left( 1 - \cos \left[ \sqrt{1 + 4 \epsilon^2} (t - \tau) \Delta E_L \right] \right) = 2 \left( 1 -  \cos \left[ (t - \tau) \Delta E_L \right] \right) \epsilon^2 + \mathcal{O}(\epsilon^4).
\end{eqnarray}

The final result for $| A_1 |^2$ reads
\begin{eqnarray}
	\label{s5.7}
	|A_1|^2 = 2 \left( 1 -  \cos \left[ (t - \tau) \Delta E_L \right] \right) A^{(1\gamma)}_{1s\, 2p} A^{(ext)}_{2p\, 2s}  A^{(2\gamma)}_{2s\,1s}
	\times 
	A^{(2\gamma)*}_{1s\,2s}A^{(ext)*}_{2s\, 2p}A^{(1\gamma)*}_{2p\, 1s}.
\end{eqnarray}
As can be seen, the leading-order term in the $\epsilon$ expansion introduces a time-dependent factor into the coefficient $C$ (or $|A_1|^2$). The natural line width of the mixed $2\tilde{s}$ state also acquires time dependence:
\begin{eqnarray}
	\label{s5.8}
	\Gamma_{2\tilde{s}}(t) \equiv \Gamma_{2s} +  2 \Gamma_{2\tilde{s}} \left( 1 - \cos \left[ (t - \tau) \Delta E_L \right] \right)  \approx \Gamma_{2s} + 2\left( 1 - \cos \left[ (t - \tau) \Delta E_L \right] \right) \left(\frac{\bm{E}}{475 \text{V/cm}}\right)^2\Gamma_{2p}.
\end{eqnarray}
Here we include the natural width of the $2s$ state, previously neglected in $\Gamma_{2\tilde{s}}$. The reason for this modification is that, for a constant uniform field, the mixed-state width is time-independent and $\Gamma_{2s}$ is negligible compared to $\Gamma_{2\tilde{s}}$. This does not hold for Eq.~(\ref{s5.8}), where the time-dependent part can vanish for particular values of $t$, leading to a zero denominator in the line profile. Alongside the time-dependent width of the mixed level, it is useful to introduce a time-dependent coefficient as
\begin{eqnarray}
	\label{s5.9}
	C(t) \equiv 2\left( 1 - \cos \left[ (t - \tau) \Delta E_L \right] \right) C.
\end{eqnarray}

As described in \cite{PhysRevLett.84.5496,Parthey,Mat,PhysRevA.59.1844}, a time delay $\tau$ is introduced to isolate the slow atoms that contribute to the spectroscopy signal. This delay corresponds to the time elapsed between the blocking of the excitation laser and the detection of the Lyman-$\alpha$ fluorescence. Taking $l$ to be the distance traveled by atoms in the $2s$ state with velocity $v$ before reaching the detection region (where they undergo prompt decay), we substitute $t = l/v$ into Eqs.~(\ref{s5.8}) and~(\ref{s5.9}). The works \cite{PhysRevLett.84.5496,Parthey,Mat} and \cite{PhysRevA.59.1844} also incorporate the second-order Doppler effect. In accordance with \cite{Kol,PhysRevA.59.1844}, we assume a Maxwellian velocity distribution:
\begin{eqnarray}
	\label{s5.10}       
	f(v) \sim v^3 e^{-(v / v_0)^2},
\end{eqnarray}
where $v_0 = \sqrt{2kT/M}$ with $M$ the hydrogen atom mass, $k$ Boltzmann's constant, and $T$ the temperature. 

Following \cite{Kol,PhysRevA.59.1844}, the velocity distribution is convolved with a Lorentzian. Then, adopting the prescription of \cite{Sob,riehle}, the frequency in the line profile is shifted according to $x \equiv \omega - \omega_0 \rightarrow \omega - \omega_0 \left( 1 - \frac{v^2}{c^2} \right) = x + \omega_0 \frac{v^2}{c^2}$, where $c$ is the speed of light and $\omega_0$ is resonant frequency. With asymmetry neglected and a Maxwellian velocity distribution assumed, the line profile takes the form:
\begin{eqnarray}
	\label{s5.11}
	\int\limits_0^{v_{\mathrm{max}}} \frac{f(v)dv}{\left[ x + \omega_0 \frac{v^2}{c^2} \right]^2 + \frac{\Gamma^2}{4}}.
\end{eqnarray}
In the general case, the upper integration limit $v_{\mathrm{max}}$ in Eq.~(\ref{s5.11}) can be set to $v_{\mathrm{max}} = \infty$. By calculating various broadening effects theoretically, a total width of $550$ Hz was obtained in \cite{Kol}, which is smaller than the experimentally observed value of $775$ Hz.

In order to incorporate both the non-adiabatic switching of the electric field and the velocity distribution of the atoms, we employ the following line profile:
\begin{eqnarray}
	\label{s5.12}
	\phi_{\tau} (x) \sim \int\limits_0^{v_{\mathrm{max}}} \frac{v^3 e^{-(v / v_0)^2} \left( 1 - \cos \left[ \left(\frac{l}{v} - \tau\right) \Delta E_L \right] \right) dv}{\left[ x + \omega_0 \frac{v^2}{c^2} \right]^2 + \frac{1}{4}\Gamma^2_{2\tilde{s}}\left(\frac{l}{v}\right)}.
\end{eqnarray}
As before, the upper integration limit may be chosen as $v_{\mathrm{max}} = \infty$ or $v_{\mathrm{max}} = l/\tau$, according to the selection of slow atoms using the time delay $\tau$. For the sake of simplicity, the asymmetry shift $\Delta(x)$ from Eq.~(\ref{s2.13}) is omitted from this line profile. Incorporating the second-order Doppler effect, the resulting expression for the frequency shift is found to be
\begin{eqnarray}
	\label{s5.13}
	\Delta(x, t) \approx \frac{4\Delta E_L^2-\Gamma_{2p}\left[\Gamma_{2p}-2\Gamma_{2\tilde{s}} (t)\right]}{4\Delta E_L^3+\Delta E_L\Gamma_{2p}\left[\Gamma_{2p}+4\Gamma_{2\tilde{s}}(t)\right]} \left[\left( x + \omega_0 \frac{v^2}{c^2} \right)^2+\frac{1}{4}\Gamma_{2\tilde{s}}(t)^2\right].
\end{eqnarray}
Equation~(\ref{s5.13}) retains only the leading-order contribution, with the factor $b/2C$ given by Eq.~(\ref{ss3.24}).

\section{Additional remarks}
\label{supp4}
According to the expression (\ref{s2.13}) the first term in $\Delta(x)$ leads to the shift $a\Gamma_{2\tilde{s}}^4/8C$ at full-width-half-maximum of the line profile. Since the estimate $\Gamma_{2\tilde{s}}\approx \left(\frac{\bm{E}}{475 \text{V/cm}}\right)^2\Gamma_{2p}$, one can find at the field strength $|\bm{E}|=10$ V/cm additional dimensionless factor $(|\bm{E}|/475)^8 \approx 3.86\times 10^{-14}$ and $\Gamma_{2p}^4/\Delta E_L^3 \sim 5.3\times 10^5$ s$^{-1}$. Further estimate can be performed considering the ratio $a\Delta E_L^3/C$: $C\approx |A_1|^2|\eta|^2$ and, therefore, $a\Delta E_L^3/C\sim 1+|A_3|^2/(|A_1|^2|\eta|^2)$.

Taking into account that the both amplitudes are proportional to the dipole photon emission matrix element, one can note
\begin{eqnarray}
\label{s4.1}
\frac{|A_3|^2}{|A_1|^2|\eta|^2} = \frac{\Delta E_L^2+\frac{1}{4}\Gamma_{2p}^2}{\frac{\bm{E}^2}{3^6}R_{2p\,2s}^2}\frac{\left|A^{(2\gamma)}_{2p\,1s}\right|^2}{\left|A^{(2\gamma)}_{2s\,1s}\right|^2},
\end{eqnarray}
where $R_{2p\,2s}^2=27$ (in atomic units) represents the radial matrix element of the dipole interaction with an external field $\bm{E}$, see, for example, \cite{SSLP_2010}. Then, for the rough estimates
\begin{eqnarray}
\label{s4.2}
\left|A^{(2\gamma)}_{2p\,1s}\right|^2 &\sim& \left(\sqrt{W_{2p\,1s}^{(E1M1)}}+\sqrt{W_{2p\,1s}^{(E1E2)}}\right)^2,
\\
\nonumber
\left|A^{(2\gamma)}_{2s\,1s}\right|^2 &\sim& W_{2s\,1s}^{(E1E1)}. 
\end{eqnarray}

In the field $|\bm{E}|\approx 1.945\times 10^{-9}$ ($10$ V/cm) and the values $W_{2s\,1s}^{(E1E1)} = 1.99\times 10^{-16}$, $W_{2p\,1s}^{(E1M1)} = 2.34\times 10^{-22}$, $W_{2p\,1s}^{(E1E2)} = 1.599\times 10^{-22}$, $\Gamma_{2p}=1.515\times 10^{-8}$, $\Delta E_L= 1.608\times 10^{-7}$ (all values are given in atomic units), one can obtain the dimensionless factor (\ref{s4.1}) as $0.73$. Thus, the squared correction
\begin{eqnarray}
\label{s4.3}
\frac{a\Gamma_{2\tilde{s}}^4}{8C} \sim 7\times 10^{-10} \text{ Hz},
\end{eqnarray}
representing the completely negligible contribution.

We would also like to emphasize once again that the results presented above were obtained for specific conditions. The angular independence of the frequency shift is closely related to the assumption of the invariance of the polarization vectors of the absorbed photons in the process. Another important condition is the opposition of the propagation directions of these photons: $\boldsymbol{\nu}_1 = - \boldsymbol{\nu}_2$. In our opinion, both assumptions are valid for experimental setups similar to those used in \cite{PhysRevLett.84.5496,Parthey,Mat}.




\end{document}